\documentclass[reprint,prb,showkeys,superscriptaddress,citeautoscript] {revtex4-2}
\usepackage{graphicx}
\usepackage{tikz}
\usepackage{braket}
\usepackage{hyperref}
\hypersetup{colorlinks=true, urlcolor= blue, citecolor=blue, linkcolor= blue, bookmarks=true, bookmarksopen=false}
\usepackage{gensymb}
\usepackage{amsmath,amssymb}
\usepackage{textcomp}
\usepackage{multirow}
\usepackage{enumitem}
\setlist{noitemsep, leftmargin=*}
\usepackage{times}
\begin{document}
\title{Origin and properties of the flat band in NbOCl$_2$ monolayer}

\author{Mohammad Ali Mohebpour}
\affiliation{Department of Physics, University of Guilan, P. O. Box 41335-1914, Rasht, Iran}

\author{Sahar Izadi Vishkayi}
\affiliation{School of Physics, Institute for Research in Fundamental Sciences (IPM), P. O. Box 19395-5531, Tehran, Iran}

\author{Valerio Vitale}
\affiliation{Dipartimento di Fisica, Università di Trieste, Strada Costiera 11, 34151 Trieste, Italy}

\author{ Nicola Seriani}
\affiliation{The Abdus Salam International Centre for Theoretical Physics (ICTP), Strada Costiera 11, I-34151 Trieste, Italy }

\author{Meysam Bagheri Tagani}
\email{m{\_}bagheri@guilan.ac.ir}
\affiliation{Department of Physics, University of Guilan, P. O. Box 41335-1914, Rasht, Iran}

\begin{abstract}
The existence of a flat band near the Fermi level can be a suitable platform for the emergence of interesting phenomena in condensed matter physics. Recently, NbOCl$_2$ monolayer has been experimentally synthesized [Nature 613 (2023) 53], which has a flat and isolated valence band. Motivated by the recent experiment, we investigate the origin of the flat band as well as the electronic, optical, photocatalytic, and magnetic properties of the monolayer by combining density functional theory and many-body quantum perturbation theory. Our results show that the flat and isolated band of this monolayer is caused by the interplay between the Peierls distortion and the electronic configuration of Nb atoms. We show that monolayers based on other elements of group 5 of the periodic table, including the V and Ta atoms, also have a flat band. The investigation of the bandwidth of the monolayer under the biaxial and uniaxial strains reveals that this material can be grown on substrates with a larger lattice constant by maintaining the flat band. Examining the material's response to the linearly polarized light not only reveals the presence of weak optical anisotropy, but also shows the existence of a bright exciton with a binding energy of about 0.94 eV. Hole doping can result in a flat band induced phase transition from semiconductor to ferromagnet. By adjusting the amount of doping, a bipolar magnetic semiconductor or a half-metal can be created. The interaction between the nearest Nb atoms is ferromagnetic, while an antiferromagnetic interaction appears between the second neighbors, which grows significantly with increasing doping. Our results demonstrate that NbOCl$_2$ monolayer has a suitable potential for spintronic applications in addition to electronic and optoelectronic applications.	
\end{abstract}

\maketitle
\section{Introduction}
Two-dimensional (2D) materials have garnered significant attention due to their exceptional properties. Among the various 2D materials that have been discovered and synthesized, graphene is the most prominent one, as it possesses remarkable electronic \cite{geim2009graphene,neto2009electre}, mechanical \cite{ovid2013mechanical,papa2017mech}, thermal \cite{balandin2011th,lin2023recent}, and optical \cite{loh2010graphene,li2019graphene} properties, which make it suitable for a wide range of applications in nanoelectronics, energy storage, sensors, and biomedicine. Graphene showed that reducing the dimensionality of a material can lead to novel physical and chemical phenomena that are not present in the bulk form. Inspired by the discovery, researchers have embarked on a quest to synthesize new 2D materials such as borophene \cite{ou2021emergence,chen2022synthesis,mozvaashi2021}, phosphorene \cite{carvalho2016phoe,batmunkh2016phoe}, h-BN \cite{naclerio2023rew,shu2022tuning}, C$_3$N \cite{yang2017c3n,zhao2023oxygen}, transition metal dichalcogenides (TMDs) \cite{yang20232d,hu2023noble}, transition metal halides (TMHs) \cite{huang2017quantum,moh2022optical}, transition metal oxide dihalides MOX$_2$ (TMOHs) \cite{fang20212d,zhao2020highly}, and MXenes \cite{wei2021advances,naguib2021ten} to offer diverse properties. These atomically thin materials exhibit incredible characteristics, such as high chemical stability, high mechanical flexibility, high carrier mobility, tunable band gap, large magnetic anisotropy, strong optical absorption, and strong excitonic effects, making them highly promising for novel devices, including high-speed transistors, flexible electronics, optoelectronics, and energy harvesting systems. Each of these 2D materials has its own unique structure and properties that can be further modified by applying strain \cite{wang2016strain,datye2022strain}, doping \cite{li2022reducing,zhao2023insight}, functionalization \cite{cui2020ad}, or defect \cite{zhao2023electrical}. In the past two decades, researchers have been actively exploring the fundamental properties and applications of 2D materials, leading to a revolution in materials science and nanotechnology.

Dispersionless band has long been considered a suitable platform for the emergence of intriguing physical phenomena. The synthesis of 2D materials and the possibility of stacking them on top of the each other have made it possible to achieve such flat bands in a limited way. At present, twisted bilayer graphene, twisted homobilater and heterostructure of some TMDs, and kagome lattices are the most famous structures with the flat band that have been synthesized. \mbox{Magic-angle} twisted bilayer graphene is a suitable platform for investigating the physics of strongly correlated electrons \cite{tarnopolsky2019origin, tilak2021flat, utama2021visualization, jiang2019charge}. Superconductivity and magnetization caused by the flat band have been confirmed experimentally in the \mbox{magic-angle} twisted bilayer graphene \cite{ yankowitz2019tuning, cao2018unconventional, lu2019superconductors, pons2020flat, song2022magic, liu2020tunable}. In addition to bilayer graphene, 2D materials with kagome lattice have recently been synthesized, which host flat bands. Han \mbox{et al.,} \cite{han2021evidence} showed that the kagome metal FeSn has a flat band with a magnetic ground state. Also, Sun \mbox{et al.,} \cite{sun2022observation} synthesized a semiconducting Nb$_3$Cl$_8$ monolayer with a breathing kagome lattice and an inversion-symmetry breaking induced band gap. Very recently, topological flat bands were observed in frustrated kagome lattice CoSn using \mbox{angle-resolved} photoemission spectroscopy \cite{kang2020topological}. This research inspired our curiosity about whether the flat band could be observed in structures other than twisted bilayers or kagome lattices.

NbOCl$_2$ monolayer has recently emerged as a captivating research subject since it has been successfully synthesized by chemical vapor transport \cite{guo2023ultrathin}. This monolayer has an orthorhombic lattice structure composed of niobium (Nb), oxygen (O), and chlorine (Cl) atoms, offering a range of intriguing properties. NbOCl$_2$ exhibits excellent flexibility, inherent chemical stability, a suitable band gap, and anisotropic optical behavior \cite{guo2023ultrathin,mortazavi2022highly}, making it an appealing platform for studying novel phenomena and developing innovative applications. This monolayer has been predicted to be a stable ferroelectric material with Curie temperature above room temperature \cite{jia2019niobium}, which makes it a promising candidate for miniaturized electronic and memory devices. The coexistence of ferroelectricity and antiferroelectricity phases makes NbOCl$_2$ monolayer exhibit easily switchable polarizations \cite{jia2019niobium}. The interplay between the transition metal and halogen atoms in NbOCl$_2$ results in a flat valence band and consequently unique electronic structure \cite{guo2023ultrathin}, which offers opportunities for tailoring its electrical and optical properties.

Despite the growing  research on NbOCl$_2$ monolayer, the origin of the flat band and possibility of the magnetization in this monolayer has not been addressed. 
Here, we investigate the origin of the flat band and the appearance of the magnetism in the monolayer using hole doping.  We employ density functional theory and \mbox{many-body} perturbative theory to comprehensively study the structural, electronic, and optical properties of NbOCl$_2$ monolayer. A spin model is presented to calculate the exchange interactions up to third nearest neighbors and magnetic anisotropy energy. We also investigate the photocatalytic performance of the monolayer for water splitting and evaluate its potential as an efficient and sustainable catalyst for this reaction. By performing the $GW-$BSE calculations, we explore the quasiparticle and excitonic properties of the monolayer. Besides, by examining the electronic band structure under different levels of hole doping, we aim to unravel the mechanism responsible for the emergence of magnetism in this monolayer. The results indicate that NbOCl$_2$ monolayer is a stable semiconductor with an indirect band gap of $2.89$~eV. The \mbox{water-splitting} performance of the monolayer could be promising under biaxial strains. The anisotropic optical coefficients suggest that the monolayer could be utilized in the development of polarization-sensitive devices such as polarizing beam splitter. The outcomes of this study not only contribute to the fundamental knowledge of NbOCl$_2$ monolayer but also pave the way to its practical applications in flexible electronic, optoelectronic, catalysis, and energy conversion devices. Importantly, this work represents the first comprehensive exploration of the magnetic properties and the emergence of magnetism in NbOCl$_2$ monolayer, shedding light on previously unexplored aspects of this intriguing monolayer.

\section{Computational Methods}
The first-principles calculations were performed based on the density functional theory (DFT) by employing the Vienna Ab-initio Simulation Package (VASP-5.4.4) \cite{kresse1996efficient}. The generalized gradient approximation (GGA) proposed by Perdew$–$Burke$–$Ernzerhof (PBE) \cite{perdew1996generalized} was chosen for describing the exchange and correlation potentials. The interactions between ion cores and valence electrons were treated with the projector$-$augmented wave (PAW) potentials \cite{gajdovs2006linear}. The \mbox{plane-wave} energy cutoff was set to be $600$~eV according to the convergence of total energy within 10$^{-3}$~eV. The Brillouin zone was integrated with a 10$\times$6$\times$1 \mbox{Monkhorst-Pack} \mbox{k-point} mesh for structural relaxation and a 16$\times$8$\times$1 \mbox{Monkhorst-Pack} \mbox{k-point} mesh for electronic calculations. The DFT$–$D2 method of Grimme \cite{grimme2006s} was applied to account for the van der Walls (vdW) interactions when performing the structural relaxation of the monolayer. The convergence criteria for energy and force were set to be 10$^{-6}$ eV and 10$^{-3}$ eV/\AA, respectively. A vacuum space of $25$~\AA~was introduced along the out-of-plane direction of the monolayer to avoid artificial interactions between adjacent images. To correct the underestimation of the band gap in the PBE method, the band structure was calculated using the Heyd$–$Scuseria$–$Ernzerhof (HSE) hybrid functional \cite{peralta2006spin}. To show the nature of the interatomic bonding, the charge transfer between atoms was determined using the Bader analysis \cite{henkelman2006fast}.

The many-body perturbative theory (MBPT) was employed to describe the many-body interactions (i.e., \mbox{electron-electron} and \mbox{electron-hole} interactions). The quasiparticle (QP) eigenvalues were determined using the state-of-the-art \mbox{single-shot} $GW$ (i.e., $G_0W_0$) approximation \cite{hybertsen1986,mohebpour2022tra}, which utilizes the \mbox{single-particle} Green’s function and screened Coulomb interactions. In this approach, the input parameters were selected based on the convergence of the band gap within 10$^{-2}$ eV. After the convergence tests, a set of $100$ conduction bands was considered in the $G_0W_0$ calculations. The number of frequency grid points was selected to be $100$. The energy cutoff for the plane wave and the response function was selected to be $600$ and $200$~eV, respectively. And, the Brillouin zone was integrated with a k-point mesh of 10$\times$6$\times$1. The excitonic and optical properties were studied by calculating the dielectric function, $\epsilon(\omega)$, from the solution of the Bethe$–$Salpeter equation (BSE) on top of the $G_0W_0$ eigenvalues (i.e., $G_0W_0-$BSE). The BSE calculation was performed based on the Tamm–Dancoff approximation (TDA) \cite{onida2002elc}, considering the $15$ lowest conduction band states and $15$ highest valence band states. The dielectric function was also computed using the random phase approximation (RPA) on top of the $G_0W_0$ eigenvalues (i.e., $G_0W_0-$RPA) to see the effects of \mbox{electron-hole} interaction on the optical response of the monolayer. The spin-orbit coupling (SOC) interaction was included in the calculations.

To check the dynamical stability of the monolayer, the phonon dispersion spectrum was calculated using the finite displacement method as implemented in the PHONOPY code \cite{togo2008}. The 2$^{nd}$-order interatomic force constants (IFCs) were obtained using a 4$\times$2$\times$1 supercell and a 4$\times$4$\times$1 k-point mesh. Moreover, to check the structural stability of the monolayer, the cohesive energy was calculated according to the formula given in Refs \cite{mohetuning,mohebpourtion}.

\section{Results and Discussion}
\subsection{Structural properties and stability}
\begin{figure*}
	\centering
	\includegraphics[width=0.92\textwidth]{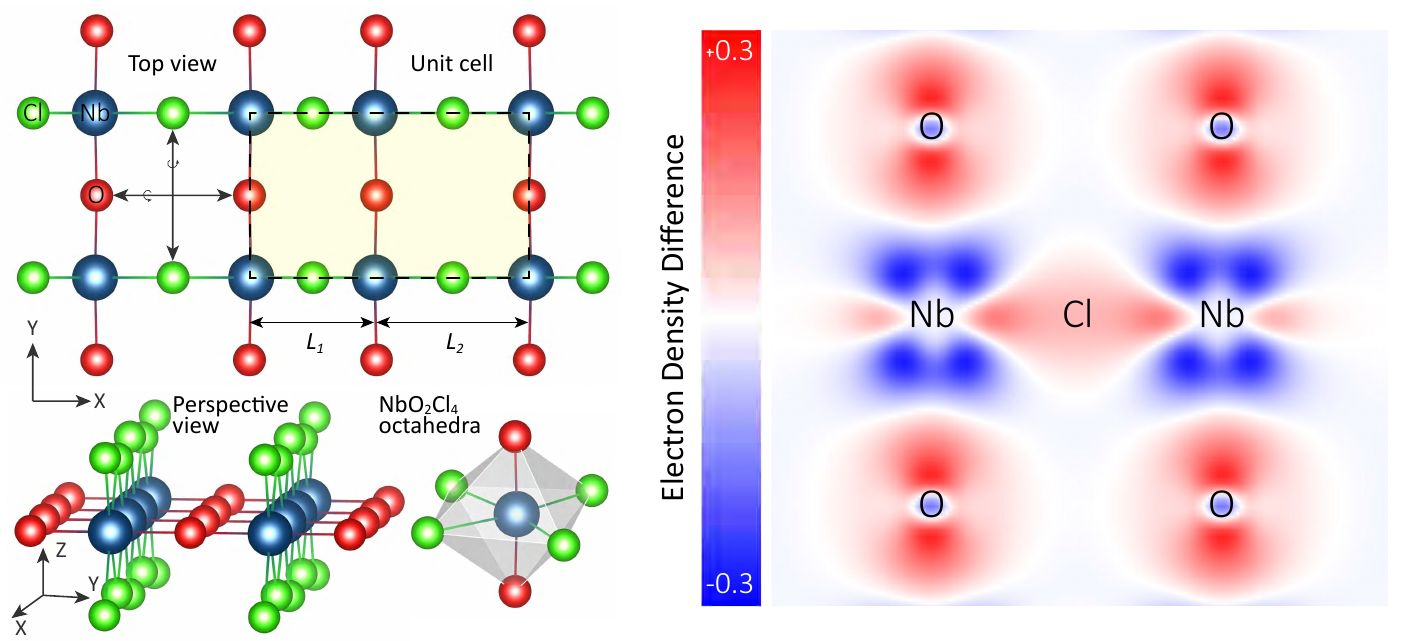}
	\caption{(Left panel) Top and perspective views of NbOCl$_2$ monolayer. The unit cell is shown by a black dashed rectangle, containing two niobium (blue balls), two oxygen (red balls), and four chlorine atoms (green balls). The mirror symmetry lines are shown by black arrows. The NbO$_2$Cl$_4$ octahedra is also displayed. (Right panel) Electron density difference profile of NbOCl$_2$ monolayer in the $x-y$ plane, showing the ionic nature of the interatomic bonding.}
	\label{struct}
\end{figure*}

We start the discussion by investigating the structural properties of the monolayer. Fig.~\ref{struct} illustrates the crystal structure of NbOCl$_2$ monolayer. As can be seen, NbOCl$_2$ monolayer has a simple orthorhombic lattice with an anisotropic planer crystal structure, where Nb atoms make bonds with Cl atoms along one planar direction and O atoms along the other one. The monolayer is composed of NbO$_2$Cl$_4$ octahedra with mixed edge- and corner-sharing connectivity. From the top view, it is clear that the Nb atoms are located at the corners of the rectangles while the O and Cl atoms are located at the center of the sides of the rectangles. From the side view, one can see that in this monolayer, the NbO layer is surrounded by two similar layers of Cl atoms. The monolayer keeps mirror symmetry under reflection in the $x-z$ and $y-z$ planes. However, it has no inversion symmetry center, which is why it exhibits strong second-order nonlinearity and piezoelectricity \cite{wu2022data}. More importantly, in this monolayer, the chain of Nb atoms shows a Peierls distortion along the $x-$axis ($L_1=3.00$~\AA~and $L_2=3.80$~\AA) that lowers the symmetry of the lattice and is responsible for the structural stability and semiconducting nature of the monolayer. Indeed, the Nb atoms form a dimerized chain along the $x-$axis of the crystal structure. Meanwhile, the O and Cl atoms form octahedral cages around the Nb atoms. Upon full relaxation, the lattice constants of the monolayer are $a=3.94$ and $b=6.73$~\AA. The ionic bond lengths of Nb$–$O and Nb$–$Cl are $2.11$ and $2.54$~\AA, respectively. Also, the thickness of the monolayer is $t=3.92$~\AA. The calculated structural parameters are in excellent agreement with previous studies \cite{mortazavi2022highly,jia2019niobium}. Stimulated scanning tunneling microscopy (STM) image is shown in Fig.~S1, in which the bright spots are the Nb atoms. The line scan along $x-$ and $y-$direction not only confirms the Peierls distortion but also demonstrates that there are three different bond lengths between Nb atoms in two directions.

The presence of a Peierls distortion in this monolayer can be a sign of the presence of a charge-density wave (CDW), so that, the ground state of the system is different from that of the single cell. In TMD monolayers, the inherent coupling of the Peierls distortion with the electronic states can be the origin of CDW \cite{chen2022ferromagnetic, cudazzo2021collective}. To investigate the existence of the CDW in NbOCl$_2$ monolayer, we also optimized $2\times1\times1$ and $2\times2\times1$ superlattices. Before the optimization, the coordinates of the atoms are slightly changed, so that in case of the CDW state, the final result becomes different from the unit cell. However, in all cases, the ground state of NbOCl$_2$ monolayer is the same. This suggests that no CDW should be present in this system.

Fig.~\ref{struct} also represents the electron density difference profile of NbOCl$_2$ monolayer. As it is clear, there is a noticeable electron accumulation around the O and Cl atoms while one can see a major electron depletion around the Nb atoms, suggesting the presence of ionic bonding in this monolayer. Generally speaking, there are two types of bonds in NbOCl$_2$ monolayer, strong Nb$–$O and weak Nb$–$Cl bonds. The Nb$–$O ionic bonds contribute to the stability and electronic properties of the monolayer. The Nb$–$Cl ionic bonds play a vital role in determining the structural integrity and chemical reactivity of the monolayer. The results are consistent with the Bader charge analysis, which indicates that $1.05$~($0.55$) charge is transmitted from the Nb atom to the O (Cl) atom, approving the ionic nature of the interatomic bonding. The charge transfer from the Nb atom to the O and Cl atoms sounds reasonable because the electronegativity of Nb ($1.60$) is much smaller than that of O ($3.44$) and Cl ($3.16$). The presence of ionic bonding in NbOCl$_2$ monolayer allows for control over the electronic and optical properties, making it favorable for certain optoelectronic applications.

It is important to note that while 2D materials with ionic bonds have unique electronic and optical properties, they may have challenges related to stability. To ensure the dynamical stability of NbOCl$_2$ monolayer, we calculated its phonon dispersion spectrum. As shown in Fig.~\ref{band}(a), all the phonon modes have real frequencies, which indicates that the lattice vibrations are well-defined and without any dynamical instability. In other words, the phonon dispersion spectrum is free of soft modes that often correspond to large displacements of atoms and can lead to phase transition or structural transformation. It is also found that the phonon modes have different dispersions along the $\Gamma-X$ and $\Gamma-Y$ directions, which is due to the anisotropy of the crystal structure. The dispersion of phonon modes along different directions can be influenced by factors such as the arrangement of atoms in the crystal structure and the strength of interatomic bonds. These factors can lead to variations in the vibrational frequencies and, consequently, different dispersions along different directions. The phonon anisotropy was already observed in the bulk counterpart \cite{guo2023ultrathin}. Moreover, we calculated the cohesive energy of the monolayer to be \mbox{$-5.52$ eV/atom}, which ensures the high structural integrity and stability of the monolayer.

\subsection{Electronic properties}
\begin{figure*}
	\centering
	\includegraphics[width=0.92\textwidth]{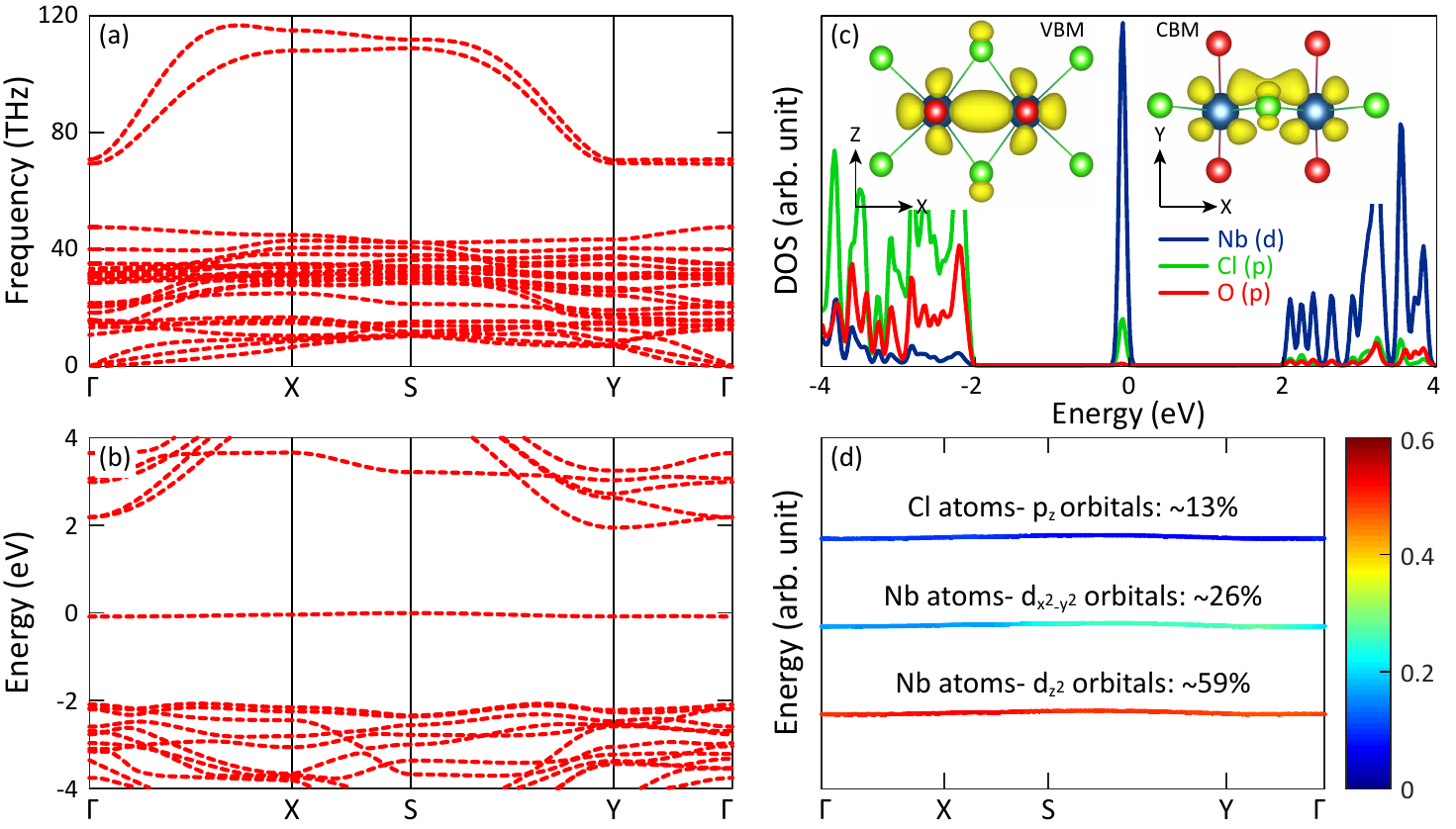}
	\caption{(a) Phonon dispersion spectrum, (b) electronic band structure, and (c) partial DOS of NbOCl$_2$ monolayer at the HSE06 level of theory. The wave function of the VBM and CBM is shown in the inset. (d) Orbital decomposed flat band, showing different orbitals contribution in the formation of the flat valence band of NbOCl$_2$ monolayer. The mean value of the orbitals contribution in the entire Brillouin zone is given.}
	\label{band}
\end{figure*}

Having established the structural properties and stability of the monolayer, we now shift our focus toward unraveling the intriguing electronic properties. Fig.~\ref{band}(b) shows the electronic band structure of NbOCl$_2$ monolayer at the HSE06 level. It is clear that NbOCl$_2$ monolayer is a semiconductor with an indirect band gap of $1.95$~eV. The valence band maximum (VBM) and conduction band minimum (CBM) of the monolayer are located at the S and Y points, respectively.
The highest valence band state is almost dispersionless through the entire Brillouin zone, such that, its bandwidth is less than $80$~meV, which suggests strongly localized holes. The flat valence band state is attributed to the structural Peierls distortion on the Nb atoms \cite{guo2023ultrathin} and is expected to enhance excitonic effects, which involve the formation of bound electron-hole pairs due to the Coulomb interactions. This flat band also enhances the electron-electron interaction in the monolayer, which can lead to exotic quantum phenomena and strongly correlated states, such as magnetism \cite{bhattacharya2023deep,li2023new}. On the contrary, the conduction band edge has almost parabolic dispersion, which allows electrons to move freely around the CBM and results in high electron mobility. Interestingly, there is an intra-band gap of $1.98$~eV between the highest valence band state and the other valence band states, which can enable new optical transitions within the valence band, such as free carrier absorption or intra-band optical pumping. Using the fitting process, the effective mass of holes is calculated to be $-8.50$ and $-12.88$~$m_0$ along the $S-X$ and $S-Y$ directions, respectively. While the effective mass of electrons is estimated to be $0.48$ and $1.78$~$m_0$ along the $Y-S$ and $Y-\Gamma$ directions, respectively. This obviously suggests that the electrons have a higher mobility than the holes and reveals the anisotropy of charge carriers along different directions.

\begin{figure}
	\centering
	\includegraphics[width=0.47\textwidth]{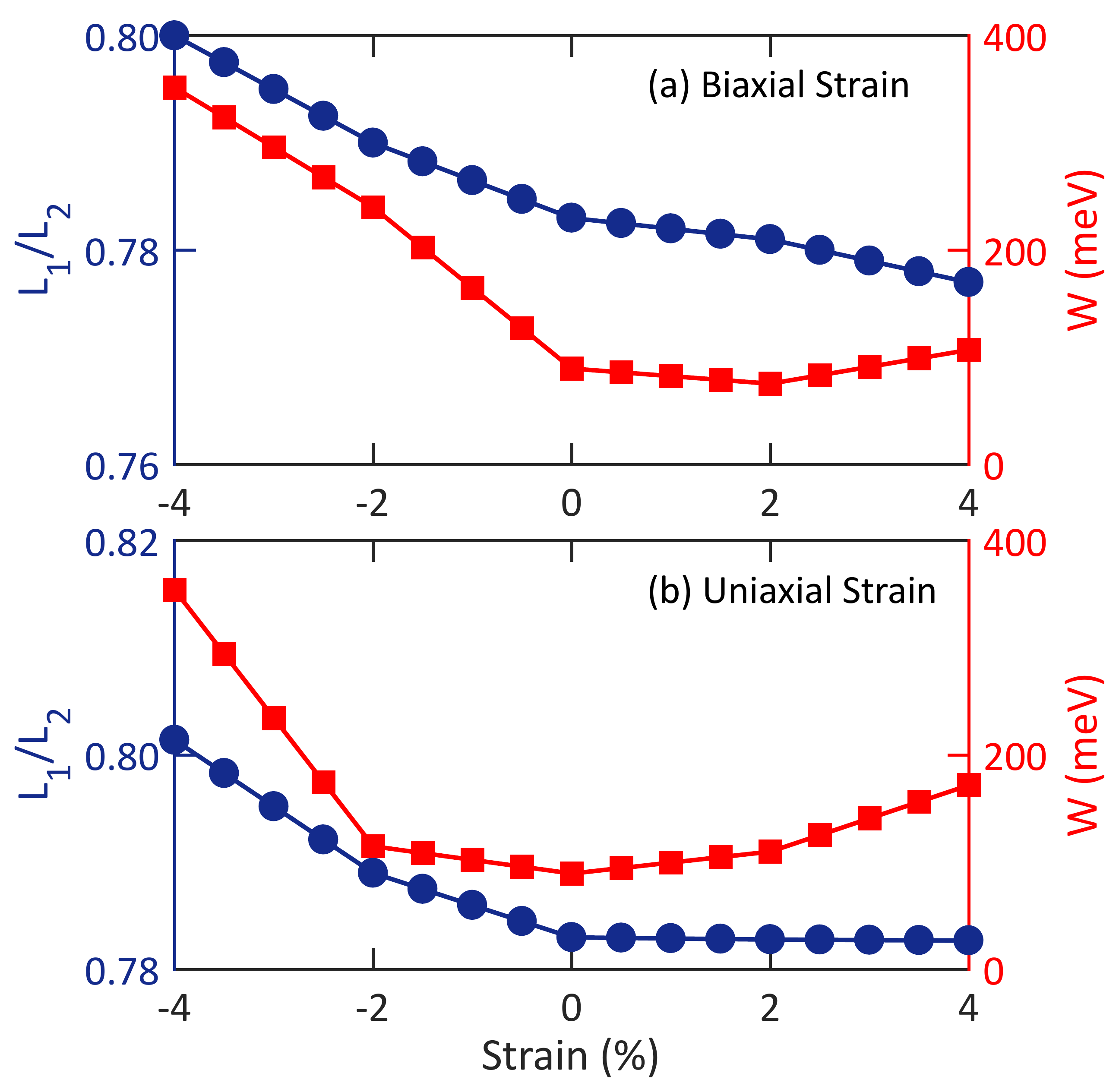}
	\caption{Variation of $L_1/L_2$ ratio and bandwidth of NbOCl$_2$ monolayer under (a) biaxial and (b) uniaxial strain along x-direction at the HSE06 level of theory.}
	\label{strain}
\end{figure}

\begin{table}
	\centering
	\normalsize
	\caption{Electronic band gap of NbOCl$_2$ monolayer under different biaxial strains at the HSE06 level of theory. The negative and positive signs signify the compression and tension, respectively.}
	\label{gap-strain}
	\begin{tabular}{lcccccccc}
		\hline
		Strain (\%)~~~~     &~~~$-4$~~~&~~~$-2$~~~&~~~$0$~~~&~~~$+2$~~~&~~~$+4$~~~\\
		\hline\hline
		Indirect band gap (eV)~~~~  &1.77 &1.83 &1.95 &2.01 &2.06 \\
		Direct band gap (eV)~~~~    &1.88 &1.92 &2.03 &2.08 &2.17 \\
		\hline
	\end{tabular}
\end{table}

Fig.~\ref{band}(c) indicates the partial DOS of NbOCl$_2$ monolayer at the HSE06 level. As can be seen, the conduction band is mostly contributed by the $d-$orbitals of the Nb atoms. Similarly, the valence band edge is fully dominated by the $d-$orbitals of the Nb atoms. However, for energies lower than $-2$~eV, the $p-$orbitals of the Cl atoms and the $p-$orbitals of the O atoms have the major contribution to the formation of the valence band states. No contribution from the O atoms is observed near the edges of the valence and conduction bands. The shape of the wave functions, as shown in the inset, indicates that the VBM and CBM are mainly formed by the $d_{z^2}$ and $d_{xy}$ orbitals of the Nb atoms, respectively. Therefore, one can say that the Nb atom has the main responsibility for the determination of the type and size of the band gap in NbOCl$_2$ monolayer. That is why the replacement of the Cl atom with Br or I atoms cannot alter the band gap of NbOX$_2$ monolayer (\mbox{X =} Cl, Br, and I) by more than $0.1$~eV \cite{mortazavi2022highly} while the substitution of the Nb atom with Zr and Hf atoms significantly increases the band gap of MOCl$_2$ monolayer (\mbox{M =} Nb, Zr, and Hf) up to $4.22$ and $4.38$~eV, respectively \cite{yang2022linear}. These are due to the structural Peierls distortion of the Nb atoms. Indeed, the Peierls distortion of the Nb atoms lowers the symmetry of the lattice and opens up a band gap of $1.95$~eV between the filled Nb$-d_{z^2}$ and empty Nb$-d_{xy}$ states, making NbOCl$_2$ monolayer a \mbox{wide-band} gap semiconductor. Removing the Peierls distortion significantly reduces the band gap of the monolayer because it restores the original symmetry and periodicity of the lattice \cite{guo2023ultrathin}. Generally speaking, the Peierls distortion play a crucial role in the formation of the band gap in NbOCl$_2$ monolayer. The formation of an energy gap between the filled and empty Nb$-d$ states was reported to be a direct consequence of the Peierls distortion of the Nb atoms in NbOI$_2$ monolayer \cite{jia2019niobium}.

One of the remarkable features of NbOCl$_2$ monolayer is the flat valence band, which is almost dispersionless through the entire Brillouin zone. The origin of this flat band is orbital hybridization. The flatness arises due to the destructive interference between different orbital contributions, leading to the cancellation of the momentum dependence of the band energy. As shown in Fig.~\ref{band}(d), the highest valence band state of the monolayer is mainly ($59$\%) composed of the Nb$-d_z^2$ orbitals, which are oriented perpendicular to the plane of the monolayer. The Nb$-d_z^2$ orbitals have a strong hybridization with the Nb$-d_{x^2-y^2}$ orbitals, which share a remarkable contribution ($26$\%). This hybridization is not a simple linear combination of the orbitals. Rather, it is a complex mixing of these orbitals with the Cl$-p_z$ orbitals ($13$\%), which are also oriented perpendicular to the plane of the monolayer. The orientation of the orbitals allows for destructive interference and results in ionic bonds between the Nb and Cl atoms, which stabilizes the energy of the highest valence band state.  It is important to note that the formation of flat bands in 2D materials is a complex phenomenon influenced by various factors, including crystal symmetry, orbital hybridization, and even electron-electron interactions. In general, the emergence of a flat band can be attributed to either topological destructive interference, known as a topological flat band, or to states associated with dangling bonds, referred to as a trivial flat band \cite{liu2014exotic}. To ascertain the nature of the flat band in this structure, we have simulated the band structure using wannier functions, as depicted in Fig S2. The comparison reveals a remarkable concordance between the Density Functional Theory (DFT) calculations and the wannier function results. The wave function corresponding to the flat band, derived from the wannier function, exhibits pronounced localization between two niobium (Nb) atoms separated by a distance  L$_2$. Consequently, it is deduced that the flat band in this material is of the trivial variety.

Although the Peierls distortion plays a vital role in the appearance of the flat band in NbOCl$_2$ monolayer, it cannot be the only reason for this phenomenon. Despite the similar crystal shape and the presence of the Peierls distortion, VOI$_2$ monolayer does not have a flat band \cite{zhang2021peierls}. MOX$_2$ monolayers (\mbox{M =} Zr and Hf; \mbox{X =} Cl, Br, and I) also have non-flat bands despite the same crystal lattice as NbOCl$_2$ monolayer and the presence of the Peierls distortion \cite{yang2022linear}. Therefore, we investigate the effect of the presence of the Peierls distortion and transition metal of group 5 of the periodic table in the emergence of the flat band.  To elucidate the origin of the flat band observed in the monolayer, we explored several scenarios. We examined the influence of Peierls distortion by analyzing a monolayer of NbOCl$_2$ in its absence. Additionally, to assess the impact of the transition metal (TM) atom's oxidation state on the flat band's formation, we investigated monolayers of ZrOCl$_2$ and MoOCl$_2$, which have TM atoms with one fewer or one additional electron compared to elements in group 5 of the periodic table. Furthermore, we replaced Nb atoms with other group 5 elements, including vanadium (V) and tantalum (Ta), to highlight the significance of group 5 elements in the flat band's emergence. 

To investigate the origin of the observed flat band in the monolayer, different scenarios are employed. To check the effect of Peierls distortion, we consider a monolayer of NbOCl$_2$ without the distortion. To study the role of the oxidation level of the TM atom in the emergence of the FB, we also consider ZrOCl$_2$ and MoOCl$_2$ monolayers having the Tm atom with one lesser or higher electron than of the element of the group 5 of the periodic table. And, we also substitute Nb atoms with other element of the group 5 including V, and Ta atoms to emphasize the role of the group 5 in the emergence of the FB. 
An analysis of the band structure of monolayer NbOCl$_2$, both with and without Peierls distortion, underscores the distortion's profound impact on the material's electronic properties, as depicted in Fig. S4 (a) and (b). In the case of a symmetric monolayer, the flat band manifests exclusively along the S-Y path of the Brillouin zone, and both the shape and band gap differ markedly from those of the actual sample. Indeed, the symmetric model behaves as a metal. Further examination of the projected density of states (DOS) indicates significant hybridization between the d-orbitals of Nb atoms at the Fermi level, prompting the transition to a metallic state. This affirms the pivotal role of Peierls distortion in shaping the distinctive electronic and optical characteristics of the monolayer NbOCl$_2$.
To explore the influence of the transition metal's (TM) oxidation state on the electronic properties of the material, we have calculated the band structures of ZrOCl$_2$ and MoOCl$_2$. As illustrated in Fig. S4 (c) and (d), ZrOCl$_2$ exhibits characteristics of a wide band gap semiconductor, whereas MoOCl$_2$ displays metallic properties. These findings corroborate that the presence of a flat band is contingent not only on Peierls distortion but also on the electronic configuration of the TM atom. The electronic configuration of Nb is [Kr]4d$^4$5s$^1$, while it is [Kr]4d$^2$5s$^2$ for Zr and [Kr]4d$^5$5s$^1$ for Mo. 
To assess the impact of TM substitution on electronic attributes, we plotted the valence band maximum (VBM) density at the $\Gamma$ point for each structure. For the optimized NbOCl$_2$ monolayer, the electron density is predominantly situated between two Nb atoms with a shorter bond length, accompanied by a discernible electron distribution on Cl atoms, aligning with the projected density of states (PDOS) results. This electron localization results in a trivial flat band. Conversely, in the symmetric monolayer, despite the electron density being present on Nb atoms, it is uniformly dispersed across all Nb atoms without localization. This contrast in electron density distribution highlights how Peierls distortion can induce a flat band.
For ZrOCl$_2$, the VBM electron density is significantly concentrated on oxygen atoms, with a notable contribution from Cl atoms. However, Zr atoms do not participate in the VBM, starkly contrasting with the Nb case. This indicates that altering the oxidation state from Nb to Zr can radically transform the electronic properties. In MoOCl$_2$, the electron density is confined to Mo atoms rather than interspersed between them. Despite a greater degree of Peierls distortion in the MoOCl$_2$ monolayer, there is no electron localization between Mo atoms with a shorter bond length, unlike the NbOCl$_2$ case.

Next, we investigate the effect of the presence of Nb atoms with partially filled $s$ and $d$ orbitals on the appearance of the flat band. To prove the importance of the existence of the element of group 5 of the periodic table in the appearance of the flat band, we replaced the Nb atoms with atoms of transition metals from the same column, i.e. the V and Ta atoms with the same number of valence electrons.  The band structure of the optimized structures is plotted in Fig.~S4. As it is clear, the flat band is observed in these two monolayers. The bandwidth for the single layer of VOCl$_2$ is equal to 90~meV, and the anisotropy coefficient of the bond between V is equal to 0.8. In the case of TaOCl$_2$ monolayer, a flat band with a width of 100 meV and a bond anisotropy coefficient of 0.78 is observed. Therefore, based on our findings and the results related to the structure of VOI$_2$ \cite{ zhang2021peierls}, it is clear that the flat band originates from the special electron arrangement of transition metals of group 5 of the periodic table, and the existence of the Peierls distortion.

To check the tunability of the band gap and electronic properties of the monolayer, we applied mechanical strain. As listed in Table~\ref{gap-strain}, the type and size of the band gap of NbOCl$_2$ monolayer are highly robust against biaxial strain. The variation of the band gap under biaxial strain from $-4$\% to $+4$\% is only $0.3$~eV. The type of the band gap is indirect. This feature is due to the ionic bonding between atoms, which makes NbO$_2$Cl$_4$ octahedra rigid and resistant to deformation.  We also investigated the dependence of the structural Peierls distortion and the bandwidth of the monolayer on biaxial strain. As shown in Fig.~\ref{strain}(a), the compressive strain increases the $L_1/L_2$ ratio, which means the reduction of the Peierls distortion and the enhancement of the crystal symmetry. Meanwhile, it increases the bandwidth of the flat valence band, which means that the flat band is a consequence of the Peierls distortion.
Therefore, one can conclude that by eliminating the structural distortion in this monolayer, the flat band disappears. Under the compressive strain, the flat band starts to widen quickly, such that, the bandwidth increases from $80$~meV at the relaxed state to about $400$~meV under the strain of $-4$\%. Overall, one can say that by applying compressive strain to the monolayer, one can change the interatomic distances and angles in the crystal, which affects the orbital hybridization, band gap, and electronic properties of the monolayer. On the contrary, the tensile strain cannot noticeably change the $L_1/L_2$ ratio. Hence, the bandwidth remains almost constant under the tensile strain.

We also applied uniaxial strain to the monolayer along the $x-$axis, the direction of the structural Peierls distortion of the Nb atoms. As shown in Fig.~\ref{strain}(b), the response of the monolayer to the uniaxial strain is similar to that to the biaxial strain. Indeed, with a slight decrease in the structural Peierls distortion, a significant increase in the bandwidth is observed. By applying uniaxial strain along the Nb$-$Nb chain direction, one can reduce the Peierls distortion and make the interatomic distances more uniform. This will increase the bandwidth of the flat band, as the electronic state become more delocalized and dispersed. By applying enough uniaxial strain, one can eliminate the Peierls distortion and restore the symmetry of the crystal structure. In this case, the flat valence band will disappear and the monolayer will become metallic, see Fig.~S2. This analysis confirms that, firstly, the structural Peierls distortion plays a fundamental role in the appearance of the flat valence band in the monolayer, and secondly, to observe the flat band, this monolayer must be grown on substrates that are subjected to an expansion strain.

By considering spin-orbit coupling (SOC) interaction in the electronic structure calculations, the CBM is split into two sub-bands, and the VBM shifts slightly upward. However, the overall shape of the band structure and the size of the band gap do not change noticeably, as shown in Fig.~S2. The size of the band gap is $1.92$~eV with SOC. This can be attributed to two main factors: the crystal symmetry of the monolayer and the moderate SOC strength of the atoms. Due to the crystal symmetry of NbOCl$_2$ monolayer, the overall effect of SOC on the electronic structure is minimal. Certainly, the mirror symmetry in NbOCl$_2$ monolayer tends to weaken the SOC contribution to the electronic structure. On the other hand, the Nb, O, and Cl atoms have moderate SOC strength, which leads to negligible effect on the electronic structure. From another point of view, one can say that SOC does not considerably affect the band gap of NbOCl$_2$ monolayer because the band gap is determined by the energy difference between the VBM and CBM, which are both composed of the $d-$orbitals of the Nb atoms.

\begin{figure}
	\centering
	\includegraphics[width=0.48\textwidth]{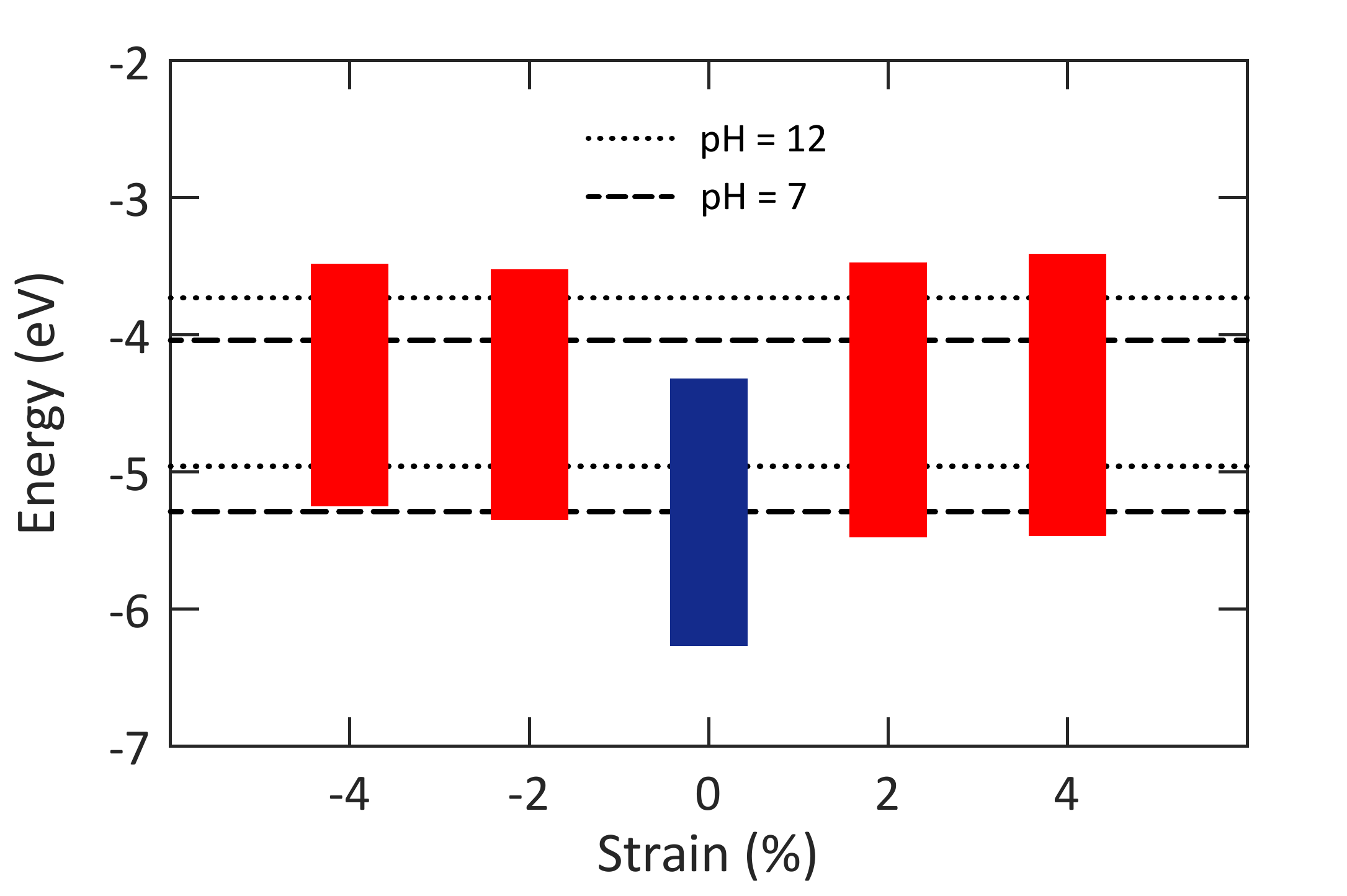}
	\caption{Valence and conduction band edge positions of NbOCl$_2$ monolayer under different biaxial strains with respect to the vacuum level at the HSE06 level of theory. The redox potentials of water splitting reaction are shown with dotted lines for \mbox{pH~$=12$} and dashed lines for \mbox{pH~$=7$}. For water splitting, at pH~$=12$~($7$), the VBM should be lower than $-4.96$~($-5.29$)~eV and the CBM should be higher than $-3.73$~($-4.04$)~eV.}
	\label{water}
\end{figure}

While the DFT$-$HSE06 calculations can provide valuable insights into the electronic properties of the monolayer, the state-of-the-art $G_0W_0$ calculation can offer a more accurate description of the electronic energy levels due to its ability to include electron-electron interaction. As shown in Fig.~S5, at the $G_0W_0+$SOC level, NbOCl$_2$ monolayer is a semiconductor with an indirect (direct) QP band gap of $2.89$ ($2.98$)~eV. This value is much larger than $0.88$ ($0.96$)~eV at the PBE$+$SOC level, showing a large QP correction ($2$~eV) and a strong electron-electron interaction. Also, it is found that the conduction band edge is parabolic while the highest valence band state is flat through the entire Brillouin zone, such that, its bandwidth is less than $0.2$~eV.

\begin{figure*}
	\centering
	\includegraphics[width=0.92\textwidth]{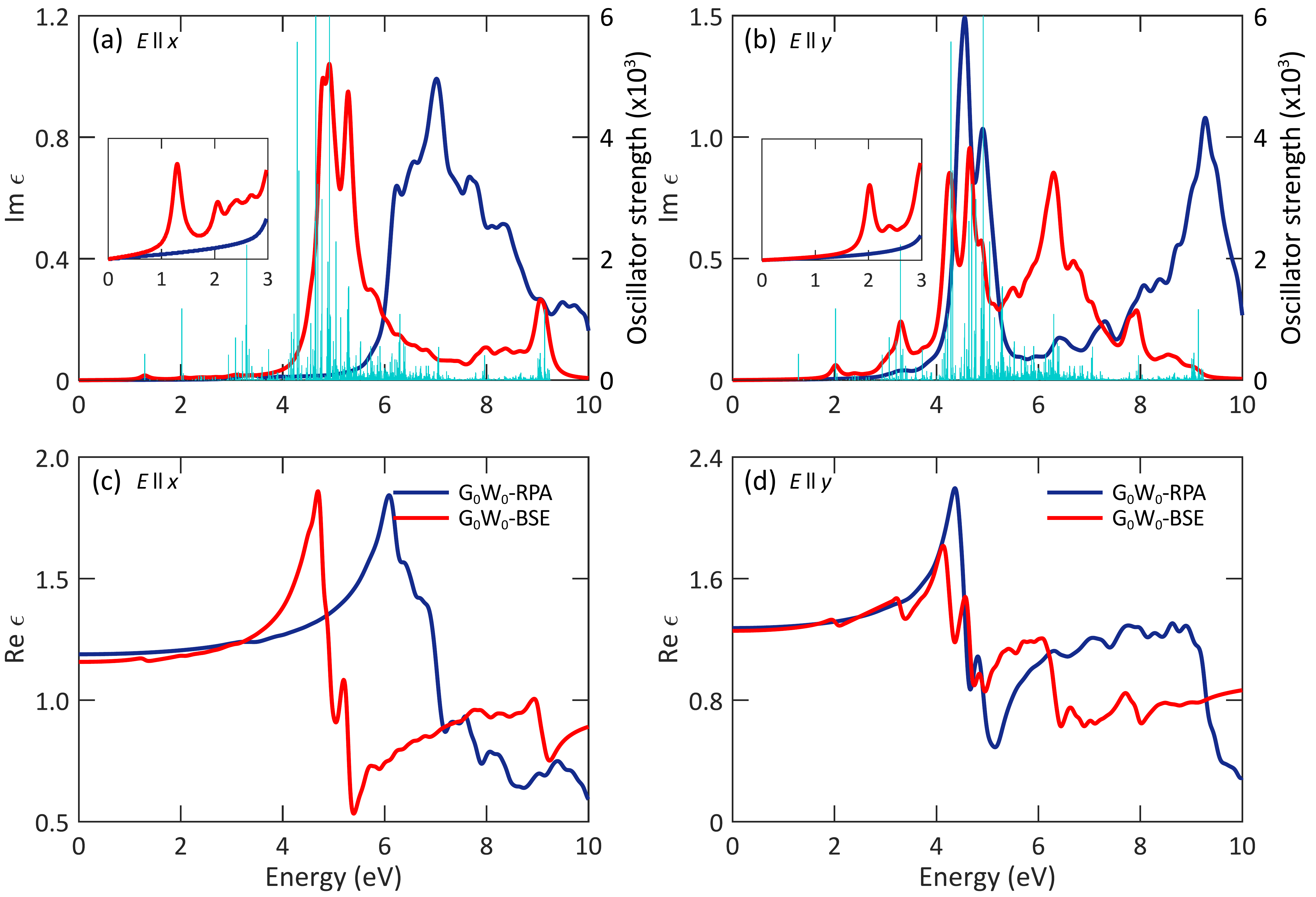}
	\caption{Imaginary (top panel) and real (bottom panel) parts of dielectric function of NbOCl$_2$ monolayer at two different levels of theory (\mbox{i.e., $G_0W_0-$RPA} and $G_0W_0-$BSE) for polarization along the $x-$direction (left panel) and $y-$direction (right panel). The insets show the same optical spectra in the range of $0$ to $3$~eV. The cyan bars indicate the oscillator strength of interband transitions.}
	\label{df}
\end{figure*}

\begin{figure}
	\centering
	\includegraphics[width=0.44\textwidth]{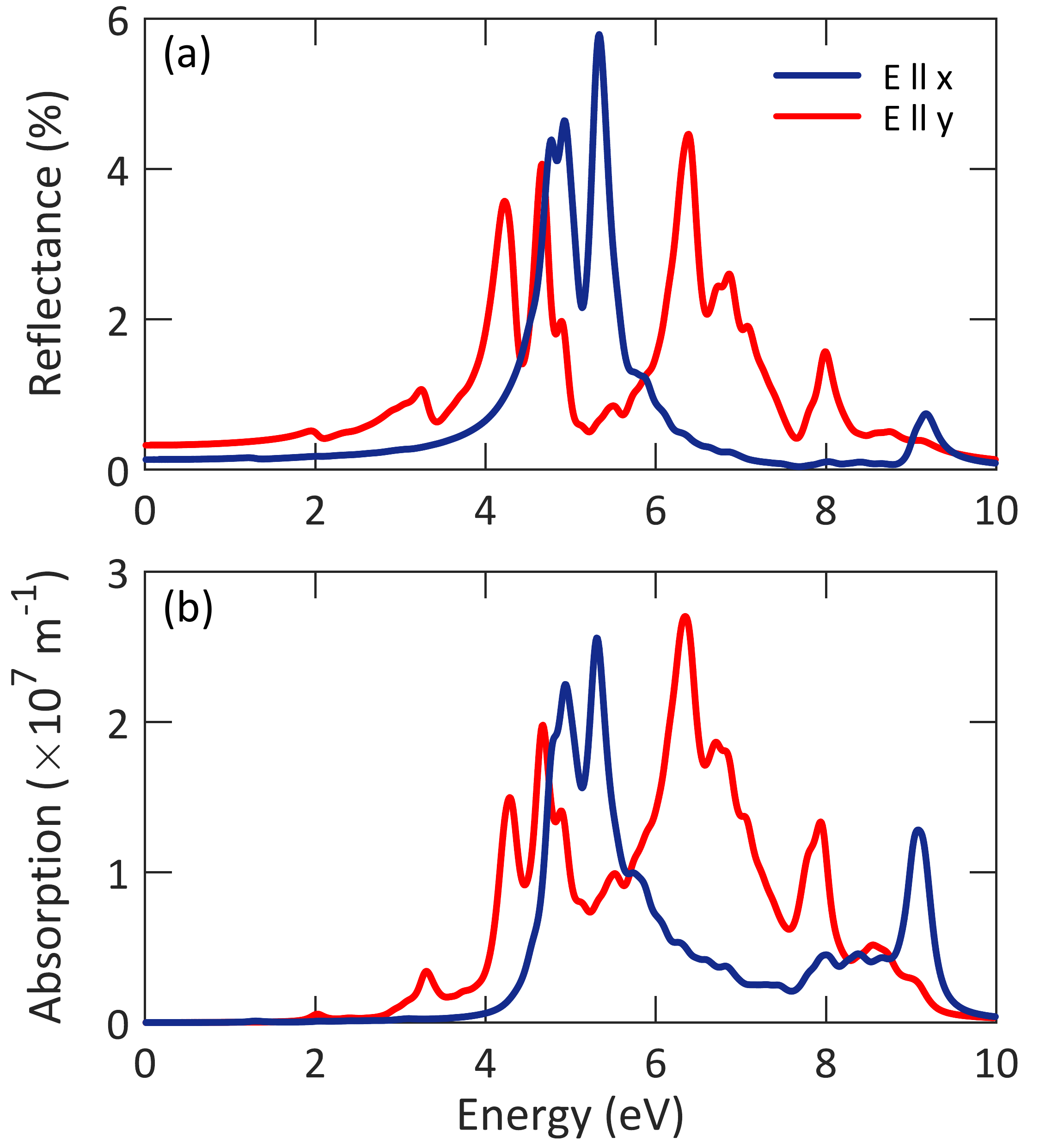}
	\caption{Optical coefficients of NbOCl$_2$ monolayer at the $G_0W_0-$BSE level for polarization along the $x$ and $y$ directions: (a) reflectance and (b) absorption.}
	\label{oc}
\end{figure}

\subsection {Photocatalytic properties}
To explore the photocatalytic performance of the monolayer for water splitting, we calculated the band edge positions of the monolayer with respect to vacuum level under different biaxial strains. As it is known, a semiconducting monolayer could be a potential photocatalyst for water splitting if the CBM energy is higher than the reduction potential of H$^+$/H$_2$ and the VBM energy is lower than the oxidation potential of O$_2$/H$_2$O. The general chemical formula for this reaction is
presented as:

\begin{equation}
\begin{split}
2H_2O+4h^+\rightarrow 4H^++O_2\\
4H^++e^-\rightarrow 2H_2
\end{split}
\end{equation}
The oxidation and reduction potentials can be tuned by changing the pH of the medium as the standard oxidation and reduction potentials of water can be written as \cite{chowdhury2017monolayer,mohebpourtion}

\begin{equation}
\begin{split}
E_{O_2/H_2O}^{ox}=-5.67 eV+0.059\times pH\\
E_{H^+/H_2}^{re}=-4.44 eV +0.059\times pH
\end{split}
\end{equation}

 As shown in Fig.~\ref{water}, for the pristine monolayer, at both \mbox{pH~$=7$} and \mbox{pH~$=12$}, the VBM energy is lower than the oxidation potential of O$_2$/H$_2$O, which means that the holes in the valence band have enough energy to oxidize water and produce oxygen. However, the CBM energy is not higher than the reduction potential of H$^+$/H$_2$, which means that the electrons in the conduction band do not have enough energy to reduce water and produce hydrogen. Therefore, the pristine monolayer is only favorable for water oxidation and O$_2$ evolution. Under the tensile strains of $+2$\% and $+4$\% and also the compressive strain of $-2$\%, both conditions are satisfied, which means that the monolayer can efficiently split water into hydrogen and oxygen using light energy without external bias. Under the compressive strain of $-4$\%, the monolayer is suitable for water splitting at \mbox{pH~$=12$}. The water-splitting performance of the monolayer could be promising under compressive strains because they reduce the structural Peierls distortion and the flatness of the highest valence band, which leads to an increase in holes mobility. Overall, one can say that applying biaxial strain to the monolayer can make its band edge positions suitable for water splitting.

\subsection{Optical properties}
In addition to electronic properties, the optical properties of NbOCl$_2$ monolayer unveil a new dimension of its fascinating behavior. Fig.~\ref{df} indicates the imaginary and real parts of the dielectric function of NbOCl$_2$ monolayer at two different levels of theory namely the $G_0W_0-$RPA (by excluding \mbox{electron-hole} interaction) and $G_0W_0-$BSE (by including \mbox{electron-hole} interaction). Obviously, the dielectric function for $x-$polarization is different from that for $y-$polarization, because the monolayer has an orthorhombic lattice that breaks the symmetry between the $x$ and $y$ directions. That is to say, NbOCl$_2$ monolayer has different electronic polarizability and local field along the $x$ and $y$ directions, which results in different responses to an applied electromagnetic field and the so-called optical anisotropy. For $x-$polarization, the imaginary part of the dielectric function at the $G_0W_0-$RPA level has a giant peak, which extends from $6$ to $10$~eV. The inclusion of \mbox{electron-hole} interaction causes a significant red shift in the imaginary part of the dielectric function. Moreover, this interaction reduces the spectral linewidth and changes the shape of the optical spectrum. For this polarization, the first peak of the optical spectrum at the $G_0W_0-$BSE level appears at $1.30$~eV, which corresponds to a strongly bound exciton with a binding energy of $1.68$~eV. However, due to the small oscillator strength ($300$) of this peak, it is a dark exciton, therefore, it cannot be called the optical gap.  The second peak of the optical spectrum lies at $2.06$~eV and has a considerable oscillator strength ($1180$), which defines a bright exciton. This exciton has a binding energy of $0.92$~eV, which is comparable to those of other well-known monolayers such as borophene ($1.13$~eV) \cite{mohebpour2022tra}, and phosphorene ($0.80$~eV) \cite{tran2014layer}. Such large binding energy indicates that the exciton can survive at room temperature and can be easily manipulated by external fields. This stability leads to improved light absorption and emission, making NbOCl$_2$ monolayer a promising candidate for optoelectronic devices such as photodetectors and light-emitting diodes (LEDs). To deepen our understanding of the optical transitions, we have illustrated the exciton wave functions for both dark and bright excitons in Fig. S7. Our analysis indicates that the dark exciton encompasses the transition from the flat band to the conduction band, whereas the bright exciton arises from the transition between the second valence band and the conduction band. The FB and the conduction bands predominantly consist of the d-orbitals from niobium (Nb) atoms. According to the optical selection rule, a parity difference of one is required between the valence and conduction bands \cite{cao2018unifying}, which accounts for the darkness of the first peak. The visualization of the exciton wave function in Fig. S7 corroborates that the electron wave function is spatially separated from the hole, extending along the Nb atoms at a bond length of  L$_2 $. Conversely, the bright exciton wave function is situated between Nb atoms with an L$_1$ bond length and is nearly localized at the hole's position. The diminished oscillator strength of the first peak is attributed to the FB's influence. Sethi et al. have demonstrated that the FB typically results in a lower oscillator strength \cite{sethi2021flat}.

 For $y-$polarization, the imaginary part of the dielectric function at the $G_0W_0-$RPA level is characterized by two primary peaks centered at $4.55$ and $9.35$~eV. The presence of electron-hole interaction results in a significant red shift in the imaginary part of the dielectric function. Besides, it reduces the intensity of the peaks. However, the main effect of electron-hole interaction for this polarization is the advent of new peaks in the optical spectrum, which are absent at the $G_0W_0-$RPA level. The first peak of the optical spectrum at the $G_0W_0-$BSE level is observed at $2.02$~eV, which corresponds to a bound exciton with a binding energy of $0.96$~eV. This value is also comparable to those of other monolayers as mentioned earlier. This peak corresponds to a bright exciton since it has a large oscillator strength ($1180$).

The peak intensity of the first bright exciton in $y-$polarization is higher than that in $x-$polarization, which means that the monolayer has a stronger optical response to light with $y-$polarization. The optical gap of the monolayer for $x$ and $y$ polarizations are slightly different because of the anisotropy of the crystal structure of the monolayer. These anisotropic optical properties suggest that the monolayer can be utilized in the development of polarization-sensitive devices such as polarizing beam splitters. Overall, based on the size of the optical gap and the shape of the optical spectra, one can conclude that NbOCl$_2$ monolayer has great potential to be used as a near-ultraviolet (UV) detector. Moreover, one can understand that the \mbox{electron-hole} interaction governs the optical behavior of the monolayer.

The effect of \mbox{electron-hole} interaction on the real part of the dielectric function is also clear. As shown in Fig.~\ref{df}(c,~d), \mbox{electron-hole} interaction significantly affects the real part of the dielectric function of the monolayer. This interaction introduces additional screening effects in the monolayer, which reduces the dielectric response of the material, leading to modifications in the shape of the real part of the dielectric function. At the $G_0W_0-$BSE level, the real part of the dielectric function exhibits a strong dependence on the energy of light, especially in the range of $4$ to $6$~eV. Due to the presence of an intra-band gap of $1.98$~eV between the Cl$-p_z$ and Nb$-d_{z^2}$ filled states in the valence band, there is no sharp peak at the onset of interband transitions. However, one can observe a giant peak at almost $4.5$~eV in both $x$ and $y$ polarizations, which is probably associated with the interband transitions from the Cl$-p_z$ states to the Nb$-d_{xy}$ states. This peak is indicative of an excitonic resonance, which arises due to strong \mbox{electron-hole} interaction within the monolayer, resulting in the formation of excitons. These excitons can significantly influence the optical properties of the monolayer. The value of the real part of the dielectric function at zero energy reflects the static dielectric constant of the monolayer, which is $1.2$ and $1.3$ for $x$ and $y$ polarizations, respectively. The difference indicates that NbOCl$_2$ monolayer has a highly anisotropic optical response.

The reflectance and absorption spectra of the monolayer at the $G_0W_0-$BSE level are plotted in Fig.~\ref{oc}. As it is clear, the monolayer shows a significant reflectance and absorption in the range of $4$ to $7$~eV for $x-$polarization, the same direction of the structural Peierls distortion. However, for $y-$polarization, the reflectance and absorption spectra are characterized by five primary peaks at different energies. This anisotropic optical response indicates that the monolayer has an inherent asymmetry, leading to different polarizabilities along different directions. Moreover, it is found that the monolayer is highly transparent in the visible light region, which shows that it is a promising candidate for transparent and flexible optoelectronic devices.

\subsection{Magnetic properties}

\begin{figure*}
	\centering
	\includegraphics[width=0.94\textwidth]{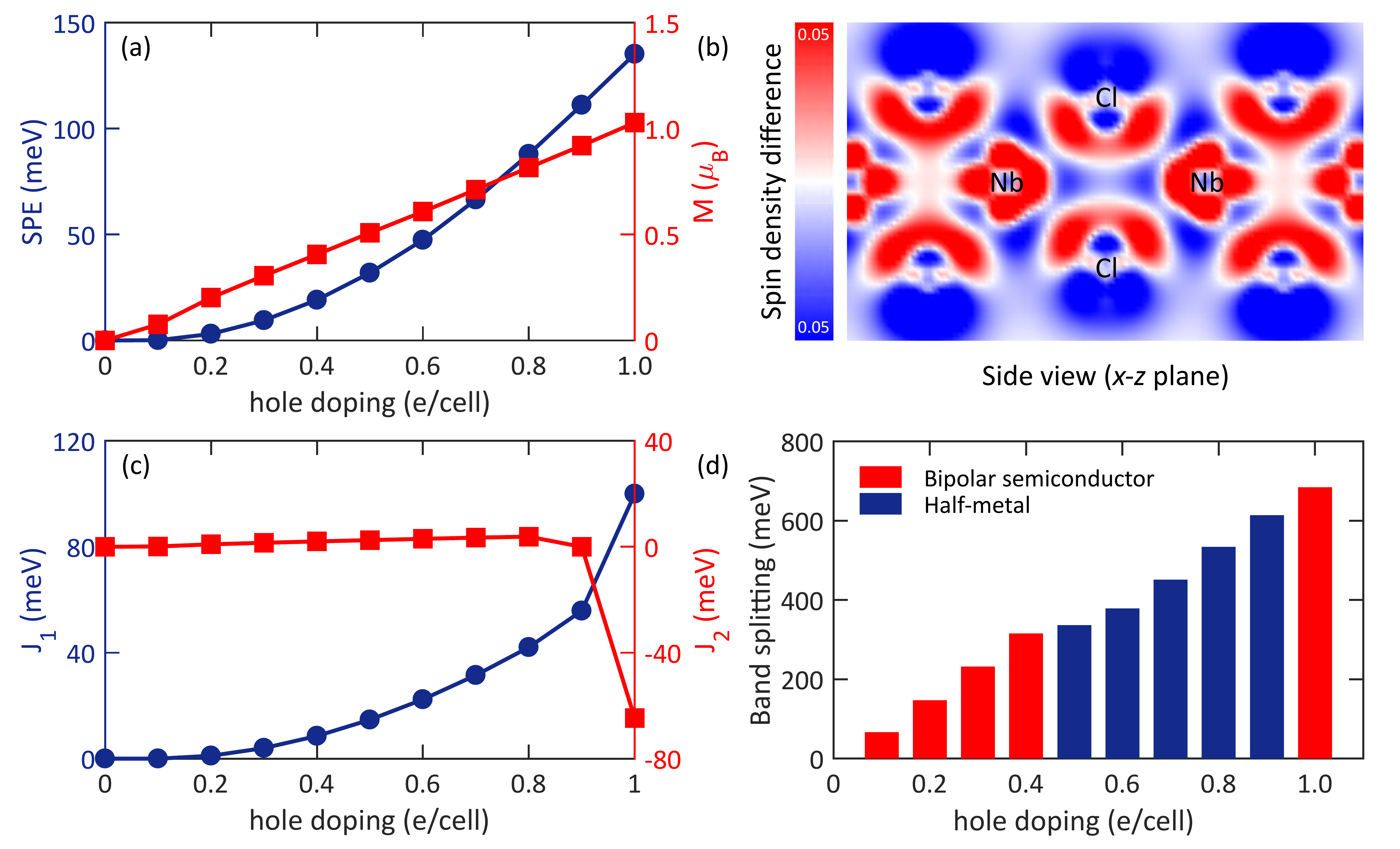}
	\caption{(a) Spin polarization energy and magnetization of NbOCl$_2$ monolayer as a function of hole doping level. (b) Spin density difference profile in the $x-z$ plane. (c) Variation of $J_1$ and $J_2$ as a function of hole doping level. (d) Band splitting as a function of hole doping level.}
	\label{spin}
\end{figure*}

A dispersionless flat valence band represents strongly localized holes and can host interesting phenomena such as magnetization and superconductivity. Considering that the valence band of NbOCl$_2$ monolayer is very close to the Fermi level, the possibility of magnetization with hole doping is probable. To investigate the emergence of magnetization in this monolayer, we calculated the electronic properties in the presence of doping. Doping has been applied by changing the total number of carriers (electrons or holes) in a unit cell and adding an opposite charge as a background in the jellium model. Recently, Deng \mbox{et al.,} \cite{deng2018gate} increased the Curie temperature of Fe$_3$GeTe$_2$ monolayer using an ionic gate. Also, Wang \mbox{et al.,} \cite{wang2018electric} used electrostatic gate doping to control the magnetization in a \mbox{few-layered} van der Waals ferromagnetic semiconducting Cr$_2$Ge$_2$Te$_6$. According to the dimensions of the unit cell, adding a hole to the monolayer is equivalent to doping of $3.8\times10^{14}$~cm$^{-2}$. We examine the spin polarization energy (EPS), which is the energy difference between the non-magnetic ground state and the spin-polarized ground state. As shown in Fig.~\ref{spin}(a), by applying doping of $0.1$~h/cell, an energy difference of $0.1$~meV appears between the two states, which indicates the appearance of magnetization. Increasing the doping level linearly increases the spin polarization energy, such that, its value reaches $135$~meV per $1$ h/cell doping.  The amount of SPE in this monolayer with less doping is much higher than the values reported for PdSe$_2$ \cite{zhang2018hole}, GaO and InO \cite{meng2020ferro}, and carbon \cite{you2019flat} monolayers. The reason for the superiority of NbOCl$_2$ monolayer compared to other monolayers is the flat valence band in the entire Brillouin zone and in the vicinity of the Fermi level, which can be easily adjusted. As expected, electron doping cannot create magnetization in the system. Mulliken analysis reveals that the magnetization observed in the monolayer is due to the local spin of the Nb atoms. As it is clear from the spin density difference plotted in Fig.~\ref{spin}(b), the Nb atoms are magnetized by hole doping. The dependence of the magnetization on the amount of doping is shown in Fig.~\ref{spin}(a). As expected, there is a linear relationship between the amount of hole doping and the magnetization appearing in the system.

The band structure of the monolayer under different hole doping concentrations is illustrated in Fig.~S6. Up to doping $0.3$~h/cell, the monolayer is a bipolar magnetic semiconductor, in which the valence and conduction bands are due to the minority and majority spins, respectively. By increasing the hole doping from $0.3$~h/cell to $0.75$~h/cell, the \mbox{semiconductor-half} metal transition occurs, such that, the minority spin band crosses the Fermi level. A further increase in hole doping turns the monolayer into a bipolar semiconductor again, but this time, the valence band is occupied by the majority spin, while the conduction band belongs to the minority spin. The presence of \mbox{half-metallic} behavior and bipolar magnetic semiconductor with a small hole doping unveils the high potential of the monolayer for spintronic applications.

We used the Heisenberg model to gain more insight into the magnetization of the hole-doped monolayer and also to investigate the \mbox{long-range} exchange interaction. The model is given by:

\begin{equation}
\label{Hee}
\begin{split}
H=-\frac{1}{2}\sum_{<i,j>}J_1S_i.S_j-\frac{1}{2}\sum_{<<i,j>>}J_2S_i.S_j
\\-\frac{1}{2}\sum_{<<<i,j>>>}J_3S_i.S_j-D\sum_{i}(S_i^z)^2
\end{split}
\end{equation}

\noindent where $J_i$ is the exchange constant between the Nb atom and its $i^{th}$ neighboring Nb atoms and $D$ denotes the magnetic anisotropy of
the structure originating from \mbox{spin-orbit} interaction. $D>0$ represents the easy axis perpendicular to the plane and $D<0$ indicates the existence of the easy plane. $S_i^z$ describes the $i^{th}$ spin of the Nb atom along the $z-$direction. Considering that there are three different bond lengths between each Nb atom and its neighboring Nb atoms, by using a 2$\times$2$\times$1 superlattice, we can investigate the spin interaction up to the third neighbor. To calculate the magnetic constants, we considered a ferromagnetic state, in which all the Nb atoms are in the same direction and four different antiferromagnetic arrangements as shown in Fig.~S7. In the antiferromagnetic state, the magnetization of the entire structure is zero. Considering the spin interaction between atoms in different configurations and using the energy obtained from \mbox{spin-polarized} calculations, the exchange energy coefficients are obtained. By using Equation~\eqref{Hee}, the energy of different spin configurations is given as:

\begin{equation}
\begin{split}
E_{FM}^z=E_0+8J_1S^2+4J_2S^2+16J_3S^2-8DS^2
\\E_{AFM1}^z=E_0+8J_1S^2+4J_2S^2-16J_3S^2-8DS^2
\\E_{AFM2}^z=E_0-8J_1S^2-4J_2S^2+16J_3S^2-8DS^2
\\E_{AFM3}^z=E_0-8J_1S^2-4J_2S^2-16J_3S^2-8DS^2
\\E_{AFM4}^z=E_0+8J_1S^2-4J_2S^2-16J_3S^2-8DS^2
\\E_{FM}^x=E_0+8J_1S^2+4J_2S^2+16J_3S^2
\end{split}
\end{equation}

\noindent where $E_0$ denotes the \mbox{spin-independent} energy part.

Fig.~\ref{spin}(c) shows the variations of the $J_1$ and $J_2$ as a function of hole doping level. It is revealed that the interaction of the nearest neighbors of the Nb atoms is ferromagnetic and the interaction strength increases uniformly with doping. In contrast, the changes in the $J_2$ with doping are very negligible up to $0.8$~h/cell doping, where the magnitude of this interaction reaches about $45$~meV, showing an antiferromagnetic interaction. The results show that there is practically no spin interaction between the third neighbors. Indeed, the spin interaction occurs between the Nb atoms that are linked with the Cl atoms, and the O atoms quench the magnetic interaction ($J_3$). Fig. S8 demonstrates the magnetic anisotropy energy (MAE) as a function of hole doping. It is revealed that the monolayer has an easy axis perpendicular to the plane of the sample. The MAE is less than that of intrinsic magnetic materials like CrI$_3$ monolayer and FeGeTe$_3$ monolayer but comparable with that of one-dimensional chain of CrI$_3$ \cite{lu2023controllable}.

We already discussed that how the uniaxial and biaxial strain can affect the Peierls distortion and bandwidth. In this part, we examine the dependence of spin polarization energy on strain. For this purpose, $\pm$ 4$\%$ biaxial strains were evaluated. The results are illustrated in Fig.~S9. Our investigation shows that this structure can still maintain its magnetic nature under the mentioned strains. A slight dependence on the pressure direction is observed in higher doping. However, the observed magnetism is robust against the pressure and can survive. This result shows the high potential of NbOCl$_2$ monolayer for application in spintronics.

Different strategies have been presented so far to create magnetism in materials \cite{lee2023possible}. Using magnetic metals such as V, Ni, and Co as doping in a bulk semiconductor is a solution that can create long-range magnetic order and leads to the emergence of diluted magnetic semiconductors. Using these magnetic metals in layered two-dimensional materials has created a new category called diluted magnetic chalcogenide semiconductors. The large magnetic anisotropy energy is responsible for the long-range magnetic order in these materials. However, the low doping concentration, its randomness, and not being adjustable with the gate are the biggest weaknesses of this class of materials. Gate tunability requires energy levels close to the Fermi level, which is difficult to reach in random doping. Recently, several intrinsic magnetic materials with a relatively low Curie temperature have been synthesized. The synthesis of these materials is difficult and they do not have good stability in environmental conditions. The two-dimensional material family of MOCl$_2$ (\mbox{M =} V, Nb, and Ta) level can be a suitable replacement for the mentioned materials due to the easier synthesis and the presence of a flat band near the Fermi level. Due to the closeness of the valence band to the Fermi level, low-hole doping using electrostatic doping induces a ferromagnetic order in this structure. The presence of magnetic anisotropy stabilizes the created order and increasing doping significantly improves the spin properties of the material. In addition, it is possible to control the transport characteristics of the material, metal or semiconductor, using electrostatic doping.

\section{Conclusion}
In summary, the structural, electronic, optical, photocatalytic, and magnetic properties of NbOCl$_2$ monolayer were investigated using density functional theory. The energy levels calculated with the $G_0W_0$ method reveals that the material is a semiconductor with an indirect band gap of 2.89~eV. The highest valence band state of the monolayer is flat and consists of d$_{z^2}$ and d$_{x^2-y^2}$ orbitals of Nb atoms, and the flat band survives under tensile strain. The monolayer can be used as a photocatalyst for water splitting under the tensile strain. We investigated the excitonic properties of the monolayer by solving the Bethe-Salpeter equation. The results reveal that this monolayer has a bright exciton with a binding energy of 0.94~eV and has a good potential to be used as a near-ultraviolet detector. The presence of a flat band can lead to the appearance of magnetism. Our investigation shows that a transition from a non-magnetic to a ferromagnetic state appears with hole doping. Using a spin model, we also investigated the exchange interaction between the first and second neighbors of the Nb atoms and reported the value of its magnetic anisotropy.

\section*{Acknowledgment}
M.B.T acknowledges the funding support by Iran National Science Foundation (INSF) under project No.4027424. 
M. B. T is thankful to the Research Council of the University
of Guilan for the partial support of this research. V. V acknowledges the funding support from the European Union’s Horizon 2020 research and innovation programme under the Marie Sk\l{}odowska-Curie grant agreement no. 101067977

\section*{Declaration of Interests}
The authors declare that they have no conflicts of interest.
\bibliographystyle{apsrev4-2}
\bibliography{APSS}

\begin{thebibliography}{78}%
\makeatletter
\providecommand \@ifxundefined [1]{%
 \@ifx{#1\undefined}
}%
\providecommand \@ifnum [1]{%
 \ifnum #1\expandafter \@firstoftwo
 \else \expandafter \@secondoftwo
 \fi
}%
\providecommand \@ifx [1]{%
 \ifx #1\expandafter \@firstoftwo
 \else \expandafter \@secondoftwo
 \fi
}%
\providecommand \natexlab [1]{#1}%
\providecommand \enquote  [1]{``#1''}%
\providecommand \bibnamefont  [1]{#1}%
\providecommand \bibfnamefont [1]{#1}%
\providecommand \citenamefont [1]{#1}%
\providecommand \href@noop [0]{\@secondoftwo}%
\providecommand \href [0]{\begingroup \@sanitize@url \@href}%
\providecommand \@href[1]{\@@startlink{#1}\@@href}%
\providecommand \@@href[1]{\endgroup#1\@@endlink}%
\providecommand \@sanitize@url [0]{\catcode `\\12\catcode `\$12\catcode
  `\&12\catcode `\#12\catcode `\^12\catcode `\_12\catcode `\%12\relax}%
\providecommand \@@startlink[1]{}%
\providecommand \@@endlink[0]{}%
\providecommand \url  [0]{\begingroup\@sanitize@url \@url }%
\providecommand \@url [1]{\endgroup\@href {#1}{\urlprefix }}%
\providecommand \urlprefix  [0]{URL }%
\providecommand \Eprint [0]{\href }%
\providecommand \doibase [0]{https://doi.org/}%
\providecommand \selectlanguage [0]{\@gobble}%
\providecommand \bibinfo  [0]{\@secondoftwo}%
\providecommand \bibfield  [0]{\@secondoftwo}%
\providecommand \translation [1]{[#1]}%
\providecommand \BibitemOpen [0]{}%
\providecommand \bibitemStop [0]{}%
\providecommand \bibitemNoStop [0]{.\EOS\space}%
\providecommand \EOS [0]{\spacefactor3000\relax}%
\providecommand \BibitemShut  [1]{\csname bibitem#1\endcsname}%
\let\auto@bib@innerbib\@empty
\bibitem [{\citenamefont {Geim}(2009)}]{geim2009graphene}%
  \BibitemOpen
  \bibfield  {author} {\bibinfo {author} {\bibfnamefont {A.~K.}\ \bibnamefont
  {Geim}},\ }\href@noop {} {\bibfield  {journal} {\bibinfo  {journal}
  {science}\ }\textbf {\bibinfo {volume} {324}},\ \bibinfo {pages} {1530}
  (\bibinfo {year} {2009})}\BibitemShut {NoStop}%
\bibitem [{\citenamefont {Neto}\ \emph {et~al.}(2009)\citenamefont {Neto},
  \citenamefont {Guinea}, \citenamefont {Peres}, \citenamefont {Novoselov},\
  and\ \citenamefont {Geim}}]{neto2009electre}%
  \BibitemOpen
  \bibfield  {author} {\bibinfo {author} {\bibfnamefont {A.~C.}\ \bibnamefont
  {Neto}}, \bibinfo {author} {\bibfnamefont {F.}~\bibnamefont {Guinea}},
  \bibinfo {author} {\bibfnamefont {N.~M.}\ \bibnamefont {Peres}}, \bibinfo
  {author} {\bibfnamefont {K.~S.}\ \bibnamefont {Novoselov}},\ and\ \bibinfo
  {author} {\bibfnamefont {A.~K.}\ \bibnamefont {Geim}},\ }\href@noop {}
  {\bibfield  {journal} {\bibinfo  {journal} {Reviews of modern physics}\
  }\textbf {\bibinfo {volume} {81}},\ \bibinfo {pages} {109} (\bibinfo {year}
  {2009})}\BibitemShut {NoStop}%
\bibitem [{\citenamefont {Ovid’Ko}(2013)}]{ovid2013mechanical}%
  \BibitemOpen
  \bibfield  {author} {\bibinfo {author} {\bibfnamefont {I.}~\bibnamefont
  {Ovid’Ko}},\ }\href@noop {} {\bibfield  {journal} {\bibinfo  {journal}
  {Rev. Adv. Mater. Sci}\ }\textbf {\bibinfo {volume} {34}},\ \bibinfo {pages}
  {1} (\bibinfo {year} {2013})}\BibitemShut {NoStop}%
\bibitem [{\citenamefont {Papageorgiou}\ \emph {et~al.}(2017)\citenamefont
  {Papageorgiou}, \citenamefont {Kinloch},\ and\ \citenamefont
  {Young}}]{papa2017mech}%
  \BibitemOpen
  \bibfield  {author} {\bibinfo {author} {\bibfnamefont {D.~G.}\ \bibnamefont
  {Papageorgiou}}, \bibinfo {author} {\bibfnamefont {I.~A.}\ \bibnamefont
  {Kinloch}},\ and\ \bibinfo {author} {\bibfnamefont {R.~J.}\ \bibnamefont
  {Young}},\ }\href@noop {} {\bibfield  {journal} {\bibinfo  {journal}
  {Progress in materials science}\ }\textbf {\bibinfo {volume} {90}},\ \bibinfo
  {pages} {75} (\bibinfo {year} {2017})}\BibitemShut {NoStop}%
\bibitem [{\citenamefont {Balandin}(2011)}]{balandin2011th}%
  \BibitemOpen
  \bibfield  {author} {\bibinfo {author} {\bibfnamefont {A.~A.}\ \bibnamefont
  {Balandin}},\ }\href@noop {} {\bibfield  {journal} {\bibinfo  {journal}
  {Nature materials}\ }\textbf {\bibinfo {volume} {10}},\ \bibinfo {pages}
  {569} (\bibinfo {year} {2011})}\BibitemShut {NoStop}%
\bibitem [{\citenamefont {Lin}\ \emph {et~al.}(2023)\citenamefont {Lin},
  \citenamefont {Jian}, \citenamefont {Bai}, \citenamefont {Li}, \citenamefont
  {Huang}, \citenamefont {Huang}, \citenamefont {Feng},\ and\ \citenamefont
  {Cheng}}]{lin2023recent}%
  \BibitemOpen
  \bibfield  {author} {\bibinfo {author} {\bibfnamefont {H.}~\bibnamefont
  {Lin}}, \bibinfo {author} {\bibfnamefont {Q.}~\bibnamefont {Jian}}, \bibinfo
  {author} {\bibfnamefont {X.}~\bibnamefont {Bai}}, \bibinfo {author}
  {\bibfnamefont {D.}~\bibnamefont {Li}}, \bibinfo {author} {\bibfnamefont
  {Z.}~\bibnamefont {Huang}}, \bibinfo {author} {\bibfnamefont
  {W.}~\bibnamefont {Huang}}, \bibinfo {author} {\bibfnamefont
  {S.}~\bibnamefont {Feng}},\ and\ \bibinfo {author} {\bibfnamefont
  {Z.}~\bibnamefont {Cheng}},\ }\href@noop {} {\bibfield  {journal} {\bibinfo
  {journal} {Applied Thermal Engineering}\ }\textbf {\bibinfo {volume} {218}},\
  \bibinfo {pages} {119176} (\bibinfo {year} {2023})}\BibitemShut {NoStop}%
\bibitem [{\citenamefont {Loh}\ \emph {et~al.}(2010)\citenamefont {Loh},
  \citenamefont {Bao}, \citenamefont {Eda},\ and\ \citenamefont
  {Chhowalla}}]{loh2010graphene}%
  \BibitemOpen
  \bibfield  {author} {\bibinfo {author} {\bibfnamefont {K.~P.}\ \bibnamefont
  {Loh}}, \bibinfo {author} {\bibfnamefont {Q.}~\bibnamefont {Bao}}, \bibinfo
  {author} {\bibfnamefont {G.}~\bibnamefont {Eda}},\ and\ \bibinfo {author}
  {\bibfnamefont {M.}~\bibnamefont {Chhowalla}},\ }\href@noop {} {\bibfield
  {journal} {\bibinfo  {journal} {Nature chemistry}\ }\textbf {\bibinfo
  {volume} {2}},\ \bibinfo {pages} {1015} (\bibinfo {year} {2010})}\BibitemShut
  {NoStop}%
\bibitem [{\citenamefont {Li}\ \emph {et~al.}(2019)\citenamefont {Li},
  \citenamefont {Zhang},\ and\ \citenamefont {Xing}}]{li2019graphene}%
  \BibitemOpen
  \bibfield  {author} {\bibinfo {author} {\bibfnamefont {Z.}~\bibnamefont
  {Li}}, \bibinfo {author} {\bibfnamefont {W.}~\bibnamefont {Zhang}},\ and\
  \bibinfo {author} {\bibfnamefont {F.}~\bibnamefont {Xing}},\ }\href@noop {}
  {\bibfield  {journal} {\bibinfo  {journal} {International journal of
  molecular sciences}\ }\textbf {\bibinfo {volume} {20}},\ \bibinfo {pages}
  {2461} (\bibinfo {year} {2019})}\BibitemShut {NoStop}%
\bibitem [{\citenamefont {Ou}\ \emph {et~al.}(2021)\citenamefont {Ou},
  \citenamefont {Wang}, \citenamefont {Yu}, \citenamefont {Liu}, \citenamefont
  {Tao}, \citenamefont {Ji},\ and\ \citenamefont {Mei}}]{ou2021emergence}%
  \BibitemOpen
  \bibfield  {author} {\bibinfo {author} {\bibfnamefont {M.}~\bibnamefont
  {Ou}}, \bibinfo {author} {\bibfnamefont {X.}~\bibnamefont {Wang}}, \bibinfo
  {author} {\bibfnamefont {L.}~\bibnamefont {Yu}}, \bibinfo {author}
  {\bibfnamefont {C.}~\bibnamefont {Liu}}, \bibinfo {author} {\bibfnamefont
  {W.}~\bibnamefont {Tao}}, \bibinfo {author} {\bibfnamefont {X.}~\bibnamefont
  {Ji}},\ and\ \bibinfo {author} {\bibfnamefont {L.}~\bibnamefont {Mei}},\
  }\href@noop {} {\bibfield  {journal} {\bibinfo  {journal} {Advanced Science}\
  }\textbf {\bibinfo {volume} {8}},\ \bibinfo {pages} {2001801} (\bibinfo
  {year} {2021})}\BibitemShut {NoStop}%
\bibitem [{\citenamefont {Chen}\ \emph
  {et~al.}(2022{\natexlab{a}})\citenamefont {Chen}, \citenamefont {Lv},
  \citenamefont {Zhang}, \citenamefont {Zhuo}, \citenamefont {Wang},
  \citenamefont {Ma}, \citenamefont {Li}, \citenamefont {Wang}, \citenamefont
  {Feng}, \citenamefont {Cheng} \emph {et~al.}}]{chen2022synthesis}%
  \BibitemOpen
  \bibfield  {author} {\bibinfo {author} {\bibfnamefont {C.}~\bibnamefont
  {Chen}}, \bibinfo {author} {\bibfnamefont {H.}~\bibnamefont {Lv}}, \bibinfo
  {author} {\bibfnamefont {P.}~\bibnamefont {Zhang}}, \bibinfo {author}
  {\bibfnamefont {Z.}~\bibnamefont {Zhuo}}, \bibinfo {author} {\bibfnamefont
  {Y.}~\bibnamefont {Wang}}, \bibinfo {author} {\bibfnamefont {C.}~\bibnamefont
  {Ma}}, \bibinfo {author} {\bibfnamefont {W.}~\bibnamefont {Li}}, \bibinfo
  {author} {\bibfnamefont {X.}~\bibnamefont {Wang}}, \bibinfo {author}
  {\bibfnamefont {B.}~\bibnamefont {Feng}}, \bibinfo {author} {\bibfnamefont
  {P.}~\bibnamefont {Cheng}}, \emph {et~al.},\ }\href@noop {} {\bibfield
  {journal} {\bibinfo  {journal} {Nature Chemistry}\ }\textbf {\bibinfo
  {volume} {14}},\ \bibinfo {pages} {25} (\bibinfo {year}
  {2022}{\natexlab{a}})}\BibitemShut {NoStop}%
\bibitem [{\citenamefont {Mozvashi}\ \emph {et~al.}(2021)\citenamefont
  {Mozvashi}, \citenamefont {Mohebpour}, \citenamefont {Vishkayi},\ and\
  \citenamefont {Tagani}}]{mozvaashi2021}%
  \BibitemOpen
  \bibfield  {author} {\bibinfo {author} {\bibfnamefont {S.~M.}\ \bibnamefont
  {Mozvashi}}, \bibinfo {author} {\bibfnamefont {M.~A.}\ \bibnamefont
  {Mohebpour}}, \bibinfo {author} {\bibfnamefont {S.~I.}\ \bibnamefont
  {Vishkayi}},\ and\ \bibinfo {author} {\bibfnamefont {M.~B.}\ \bibnamefont
  {Tagani}},\ }\href@noop {} {\bibfield  {journal} {\bibinfo  {journal}
  {Scientific Reports}\ }\textbf {\bibinfo {volume} {11}},\ \bibinfo {pages}
  {7547} (\bibinfo {year} {2021})}\BibitemShut {NoStop}%
\bibitem [{\citenamefont {Carvalho}\ \emph {et~al.}(2016)\citenamefont
  {Carvalho}, \citenamefont {Wang}, \citenamefont {Zhu}, \citenamefont {Rodin},
  \citenamefont {Su},\ and\ \citenamefont {Castro~Neto}}]{carvalho2016phoe}%
  \BibitemOpen
  \bibfield  {author} {\bibinfo {author} {\bibfnamefont {A.}~\bibnamefont
  {Carvalho}}, \bibinfo {author} {\bibfnamefont {M.}~\bibnamefont {Wang}},
  \bibinfo {author} {\bibfnamefont {X.}~\bibnamefont {Zhu}}, \bibinfo {author}
  {\bibfnamefont {A.~S.}\ \bibnamefont {Rodin}}, \bibinfo {author}
  {\bibfnamefont {H.}~\bibnamefont {Su}},\ and\ \bibinfo {author}
  {\bibfnamefont {A.~H.}\ \bibnamefont {Castro~Neto}},\ }\href@noop {}
  {\bibfield  {journal} {\bibinfo  {journal} {Nature Reviews Materials}\
  }\textbf {\bibinfo {volume} {1}},\ \bibinfo {pages} {1} (\bibinfo {year}
  {2016})}\BibitemShut {NoStop}%
\bibitem [{\citenamefont {Batmunkh}\ \emph {et~al.}(2016)\citenamefont
  {Batmunkh}, \citenamefont {Bat-Erdene},\ and\ \citenamefont
  {Shapter}}]{batmunkh2016phoe}%
  \BibitemOpen
  \bibfield  {author} {\bibinfo {author} {\bibfnamefont {M.}~\bibnamefont
  {Batmunkh}}, \bibinfo {author} {\bibfnamefont {M.}~\bibnamefont
  {Bat-Erdene}},\ and\ \bibinfo {author} {\bibfnamefont {J.~G.}\ \bibnamefont
  {Shapter}},\ }\href@noop {} {\bibfield  {journal} {\bibinfo  {journal}
  {Advanced Materials}\ }\textbf {\bibinfo {volume} {28}},\ \bibinfo {pages}
  {8586} (\bibinfo {year} {2016})}\BibitemShut {NoStop}%
\bibitem [{\citenamefont {Naclerio}\ and\ \citenamefont
  {Kidambi}(2023)}]{naclerio2023rew}%
  \BibitemOpen
  \bibfield  {author} {\bibinfo {author} {\bibfnamefont {A.~E.}\ \bibnamefont
  {Naclerio}}\ and\ \bibinfo {author} {\bibfnamefont {P.~R.}\ \bibnamefont
  {Kidambi}},\ }\href@noop {} {\bibfield  {journal} {\bibinfo  {journal}
  {Advanced Materials}\ }\textbf {\bibinfo {volume} {35}},\ \bibinfo {pages}
  {2207374} (\bibinfo {year} {2023})}\BibitemShut {NoStop}%
\bibitem [{\citenamefont {Shu}\ and\ \citenamefont
  {Liu}(2022)}]{shu2022tuning}%
  \BibitemOpen
  \bibfield  {author} {\bibinfo {author} {\bibfnamefont {H.}~\bibnamefont
  {Shu}}\ and\ \bibinfo {author} {\bibfnamefont {X.}~\bibnamefont {Liu}},\
  }\href@noop {} {\bibfield  {journal} {\bibinfo  {journal} {Applied Surface
  Science}\ }\textbf {\bibinfo {volume} {605}},\ \bibinfo {pages} {154591}
  (\bibinfo {year} {2022})}\BibitemShut {NoStop}%
\bibitem [{\citenamefont {Yang}\ \emph {et~al.}(2017)\citenamefont {Yang},
  \citenamefont {Li}, \citenamefont {Ye}, \citenamefont {Wang}, \citenamefont
  {Tian}, \citenamefont {Zhu}, \citenamefont {He}, \citenamefont {Ding},
  \citenamefont {Xie}, \citenamefont {Liu} \emph {et~al.}}]{yang2017c3n}%
  \BibitemOpen
  \bibfield  {author} {\bibinfo {author} {\bibfnamefont {S.}~\bibnamefont
  {Yang}}, \bibinfo {author} {\bibfnamefont {W.}~\bibnamefont {Li}}, \bibinfo
  {author} {\bibfnamefont {C.}~\bibnamefont {Ye}}, \bibinfo {author}
  {\bibfnamefont {G.}~\bibnamefont {Wang}}, \bibinfo {author} {\bibfnamefont
  {H.}~\bibnamefont {Tian}}, \bibinfo {author} {\bibfnamefont {C.}~\bibnamefont
  {Zhu}}, \bibinfo {author} {\bibfnamefont {P.}~\bibnamefont {He}}, \bibinfo
  {author} {\bibfnamefont {G.}~\bibnamefont {Ding}}, \bibinfo {author}
  {\bibfnamefont {X.}~\bibnamefont {Xie}}, \bibinfo {author} {\bibfnamefont
  {Y.}~\bibnamefont {Liu}}, \emph {et~al.},\ }\href@noop {} {\bibfield
  {journal} {\bibinfo  {journal} {Advanced Materials}\ }\textbf {\bibinfo
  {volume} {29}},\ \bibinfo {pages} {1605625} (\bibinfo {year}
  {2017})}\BibitemShut {NoStop}%
\bibitem [{\citenamefont {Zhao}\ \emph
  {et~al.}(2023{\natexlab{a}})\citenamefont {Zhao}, \citenamefont {Luo},
  \citenamefont {Huang}, \citenamefont {Yan}, \citenamefont {Jia},
  \citenamefont {Pei},\ and\ \citenamefont {Tu}}]{zhao2023oxygen}%
  \BibitemOpen
  \bibfield  {author} {\bibinfo {author} {\bibfnamefont {L.}~\bibnamefont
  {Zhao}}, \bibinfo {author} {\bibfnamefont {W.}~\bibnamefont {Luo}}, \bibinfo
  {author} {\bibfnamefont {Z.}~\bibnamefont {Huang}}, \bibinfo {author}
  {\bibfnamefont {Z.}~\bibnamefont {Yan}}, \bibinfo {author} {\bibfnamefont
  {H.}~\bibnamefont {Jia}}, \bibinfo {author} {\bibfnamefont {W.}~\bibnamefont
  {Pei}},\ and\ \bibinfo {author} {\bibfnamefont {Y.}~\bibnamefont {Tu}},\
  }\href@noop {} {\bibfield  {journal} {\bibinfo  {journal} {Applied Surface
  Science}\ }\textbf {\bibinfo {volume} {613}},\ \bibinfo {pages} {155912}
  (\bibinfo {year} {2023}{\natexlab{a}})}\BibitemShut {NoStop}%
\bibitem [{\citenamefont {Yang}\ \emph {et~al.}(2023)\citenamefont {Yang},
  \citenamefont {Fan}, \citenamefont {Zhang}, \citenamefont {Mei},
  \citenamefont {Zhu}, \citenamefont {Qin}, \citenamefont {Hu}, \citenamefont
  {Chen}, \citenamefont {Hau~Ng}, \citenamefont {Voiry} \emph
  {et~al.}}]{yang20232d}%
  \BibitemOpen
  \bibfield  {author} {\bibinfo {author} {\bibfnamefont {R.}~\bibnamefont
  {Yang}}, \bibinfo {author} {\bibfnamefont {Y.}~\bibnamefont {Fan}}, \bibinfo
  {author} {\bibfnamefont {Y.}~\bibnamefont {Zhang}}, \bibinfo {author}
  {\bibfnamefont {L.}~\bibnamefont {Mei}}, \bibinfo {author} {\bibfnamefont
  {R.}~\bibnamefont {Zhu}}, \bibinfo {author} {\bibfnamefont {J.}~\bibnamefont
  {Qin}}, \bibinfo {author} {\bibfnamefont {J.}~\bibnamefont {Hu}}, \bibinfo
  {author} {\bibfnamefont {Z.}~\bibnamefont {Chen}}, \bibinfo {author}
  {\bibfnamefont {Y.}~\bibnamefont {Hau~Ng}}, \bibinfo {author} {\bibfnamefont
  {D.}~\bibnamefont {Voiry}}, \emph {et~al.},\ }\href@noop {} {\bibfield
  {journal} {\bibinfo  {journal} {Angewandte Chemie International Edition}\
  }\textbf {\bibinfo {volume} {62}},\ \bibinfo {pages} {e202218016} (\bibinfo
  {year} {2023})}\BibitemShut {NoStop}%
\bibitem [{\citenamefont {Hu}\ \emph {et~al.}(2023)\citenamefont {Hu},
  \citenamefont {Zheng}, \citenamefont {Fan}, \citenamefont {Zhang},\ and\
  \citenamefont {Liu}}]{hu2023noble}%
  \BibitemOpen
  \bibfield  {author} {\bibinfo {author} {\bibfnamefont {Y.}~\bibnamefont
  {Hu}}, \bibinfo {author} {\bibfnamefont {W.}~\bibnamefont {Zheng}}, \bibinfo
  {author} {\bibfnamefont {S.}~\bibnamefont {Fan}}, \bibinfo {author}
  {\bibfnamefont {J.}~\bibnamefont {Zhang}},\ and\ \bibinfo {author}
  {\bibfnamefont {X.}~\bibnamefont {Liu}},\ }\href@noop {} {\bibfield
  {journal} {\bibinfo  {journal} {Applied Physics Reviews}\ }\textbf {\bibinfo
  {volume} {10}} (\bibinfo {year} {2023})}\BibitemShut {NoStop}%
\bibitem [{\citenamefont {Huang}\ \emph {et~al.}(2017)\citenamefont {Huang},
  \citenamefont {Zhou}, \citenamefont {Wu}, \citenamefont {Deng}, \citenamefont
  {Jena},\ and\ \citenamefont {Kan}}]{huang2017quantum}%
  \BibitemOpen
  \bibfield  {author} {\bibinfo {author} {\bibfnamefont {C.}~\bibnamefont
  {Huang}}, \bibinfo {author} {\bibfnamefont {J.}~\bibnamefont {Zhou}},
  \bibinfo {author} {\bibfnamefont {H.}~\bibnamefont {Wu}}, \bibinfo {author}
  {\bibfnamefont {K.}~\bibnamefont {Deng}}, \bibinfo {author} {\bibfnamefont
  {P.}~\bibnamefont {Jena}},\ and\ \bibinfo {author} {\bibfnamefont
  {E.}~\bibnamefont {Kan}},\ }\href@noop {} {\bibfield  {journal} {\bibinfo
  {journal} {Physical Review B}\ }\textbf {\bibinfo {volume} {95}},\ \bibinfo
  {pages} {045113} (\bibinfo {year} {2017})}\BibitemShut {NoStop}%
\bibitem [{\citenamefont {Mohebpour}\ \emph
  {et~al.}(2022{\natexlab{a}})\citenamefont {Mohebpour}, \citenamefont
  {Mortazavi}, \citenamefont {Zhuang},\ and\ \citenamefont
  {Tagani}}]{moh2022optical}%
  \BibitemOpen
  \bibfield  {author} {\bibinfo {author} {\bibfnamefont {M.~A.}\ \bibnamefont
  {Mohebpour}}, \bibinfo {author} {\bibfnamefont {B.}~\bibnamefont
  {Mortazavi}}, \bibinfo {author} {\bibfnamefont {X.}~\bibnamefont {Zhuang}},\
  and\ \bibinfo {author} {\bibfnamefont {M.~B.}\ \bibnamefont {Tagani}},\
  }\href@noop {} {\bibfield  {journal} {\bibinfo  {journal} {Physical Review
  B}\ }\textbf {\bibinfo {volume} {106}},\ \bibinfo {pages} {125405} (\bibinfo
  {year} {2022}{\natexlab{a}})}\BibitemShut {NoStop}%
\bibitem [{\citenamefont {Fang}\ \emph {et~al.}(2021)\citenamefont {Fang},
  \citenamefont {Wang}, \citenamefont {Wang}, \citenamefont {Zhai},\ and\
  \citenamefont {Huang}}]{fang20212d}%
  \BibitemOpen
  \bibfield  {author} {\bibinfo {author} {\bibfnamefont {Y.}~\bibnamefont
  {Fang}}, \bibinfo {author} {\bibfnamefont {F.}~\bibnamefont {Wang}}, \bibinfo
  {author} {\bibfnamefont {R.}~\bibnamefont {Wang}}, \bibinfo {author}
  {\bibfnamefont {T.}~\bibnamefont {Zhai}},\ and\ \bibinfo {author}
  {\bibfnamefont {F.}~\bibnamefont {Huang}},\ }\href@noop {} {\bibfield
  {journal} {\bibinfo  {journal} {Advanced Materials}\ }\textbf {\bibinfo
  {volume} {33}},\ \bibinfo {pages} {2101505} (\bibinfo {year}
  {2021})}\BibitemShut {NoStop}%
\bibitem [{\citenamefont {Zhao}\ \emph {et~al.}(2020)\citenamefont {Zhao},
  \citenamefont {Wu}, \citenamefont {Zhu}, \citenamefont {Lu}, \citenamefont
  {Xiang},\ and\ \citenamefont {Yang}}]{zhao2020highly}%
  \BibitemOpen
  \bibfield  {author} {\bibinfo {author} {\bibfnamefont {J.}~\bibnamefont
  {Zhao}}, \bibinfo {author} {\bibfnamefont {W.}~\bibnamefont {Wu}}, \bibinfo
  {author} {\bibfnamefont {J.}~\bibnamefont {Zhu}}, \bibinfo {author}
  {\bibfnamefont {Y.}~\bibnamefont {Lu}}, \bibinfo {author} {\bibfnamefont
  {B.}~\bibnamefont {Xiang}},\ and\ \bibinfo {author} {\bibfnamefont {S.~A.}\
  \bibnamefont {Yang}},\ }\href@noop {} {\bibfield  {journal} {\bibinfo
  {journal} {Physical Review B}\ }\textbf {\bibinfo {volume} {102}},\ \bibinfo
  {pages} {245419} (\bibinfo {year} {2020})}\BibitemShut {NoStop}%
\bibitem [{\citenamefont {Wei}\ \emph {et~al.}(2021)\citenamefont {Wei},
  \citenamefont {Zhang}, \citenamefont {Soomro}, \citenamefont {Zhu},\ and\
  \citenamefont {Xu}}]{wei2021advances}%
  \BibitemOpen
  \bibfield  {author} {\bibinfo {author} {\bibfnamefont {Y.}~\bibnamefont
  {Wei}}, \bibinfo {author} {\bibfnamefont {P.}~\bibnamefont {Zhang}}, \bibinfo
  {author} {\bibfnamefont {R.~A.}\ \bibnamefont {Soomro}}, \bibinfo {author}
  {\bibfnamefont {Q.}~\bibnamefont {Zhu}},\ and\ \bibinfo {author}
  {\bibfnamefont {B.}~\bibnamefont {Xu}},\ }\href@noop {} {\bibfield  {journal}
  {\bibinfo  {journal} {Advanced materials}\ }\textbf {\bibinfo {volume}
  {33}},\ \bibinfo {pages} {2103148} (\bibinfo {year} {2021})}\BibitemShut
  {NoStop}%
\bibitem [{\citenamefont {Naguib}\ \emph {et~al.}(2021)\citenamefont {Naguib},
  \citenamefont {Barsoum},\ and\ \citenamefont {Gogotsi}}]{naguib2021ten}%
  \BibitemOpen
  \bibfield  {author} {\bibinfo {author} {\bibfnamefont {M.}~\bibnamefont
  {Naguib}}, \bibinfo {author} {\bibfnamefont {M.~W.}\ \bibnamefont
  {Barsoum}},\ and\ \bibinfo {author} {\bibfnamefont {Y.}~\bibnamefont
  {Gogotsi}},\ }\href@noop {} {\bibfield  {journal} {\bibinfo  {journal}
  {Advanced Materials}\ }\textbf {\bibinfo {volume} {33}},\ \bibinfo {pages}
  {2103393} (\bibinfo {year} {2021})}\BibitemShut {NoStop}%
\bibitem [{\citenamefont {Wang}\ \emph {et~al.}(2016)\citenamefont {Wang},
  \citenamefont {Wang}, \citenamefont {Lu}, \citenamefont {Jiang},\ and\
  \citenamefont {Yang}}]{wang2016strain}%
  \BibitemOpen
  \bibfield  {author} {\bibinfo {author} {\bibfnamefont {Y.}~\bibnamefont
  {Wang}}, \bibinfo {author} {\bibfnamefont {S.-S.}\ \bibnamefont {Wang}},
  \bibinfo {author} {\bibfnamefont {Y.}~\bibnamefont {Lu}}, \bibinfo {author}
  {\bibfnamefont {J.}~\bibnamefont {Jiang}},\ and\ \bibinfo {author}
  {\bibfnamefont {S.~A.}\ \bibnamefont {Yang}},\ }\href@noop {} {\bibfield
  {journal} {\bibinfo  {journal} {Nano letters}\ }\textbf {\bibinfo {volume}
  {16}},\ \bibinfo {pages} {4576} (\bibinfo {year} {2016})}\BibitemShut
  {NoStop}%
\bibitem [{\citenamefont {Datye}\ \emph {et~al.}(2022)\citenamefont {Datye},
  \citenamefont {Daus}, \citenamefont {Grady}, \citenamefont {Brenner},
  \citenamefont {Vaziri},\ and\ \citenamefont {Pop}}]{datye2022strain}%
  \BibitemOpen
  \bibfield  {author} {\bibinfo {author} {\bibfnamefont {I.~M.}\ \bibnamefont
  {Datye}}, \bibinfo {author} {\bibfnamefont {A.}~\bibnamefont {Daus}},
  \bibinfo {author} {\bibfnamefont {R.~W.}\ \bibnamefont {Grady}}, \bibinfo
  {author} {\bibfnamefont {K.}~\bibnamefont {Brenner}}, \bibinfo {author}
  {\bibfnamefont {S.}~\bibnamefont {Vaziri}},\ and\ \bibinfo {author}
  {\bibfnamefont {E.}~\bibnamefont {Pop}},\ }\href@noop {} {\bibfield
  {journal} {\bibinfo  {journal} {Nano Letters}\ }\textbf {\bibinfo {volume}
  {22}},\ \bibinfo {pages} {8052} (\bibinfo {year} {2022})}\BibitemShut
  {NoStop}%
\bibitem [{\citenamefont {Li}\ \emph {et~al.}(2022)\citenamefont {Li},
  \citenamefont {Cheng}, \citenamefont {Wang}, \citenamefont {Du},
  \citenamefont {Song}, \citenamefont {He},\ and\ \citenamefont
  {Shi}}]{li2022reducing}%
  \BibitemOpen
  \bibfield  {author} {\bibinfo {author} {\bibfnamefont {H.}~\bibnamefont
  {Li}}, \bibinfo {author} {\bibfnamefont {M.}~\bibnamefont {Cheng}}, \bibinfo
  {author} {\bibfnamefont {P.}~\bibnamefont {Wang}}, \bibinfo {author}
  {\bibfnamefont {R.}~\bibnamefont {Du}}, \bibinfo {author} {\bibfnamefont
  {L.}~\bibnamefont {Song}}, \bibinfo {author} {\bibfnamefont {J.}~\bibnamefont
  {He}},\ and\ \bibinfo {author} {\bibfnamefont {J.}~\bibnamefont {Shi}},\
  }\href@noop {} {\bibfield  {journal} {\bibinfo  {journal} {Advanced
  Materials}\ }\textbf {\bibinfo {volume} {34}},\ \bibinfo {pages} {2200885}
  (\bibinfo {year} {2022})}\BibitemShut {NoStop}%
\bibitem [{\citenamefont {Zhao}\ \emph
  {et~al.}(2023{\natexlab{b}})\citenamefont {Zhao}, \citenamefont {Wan},
  \citenamefont {Qian},\ and\ \citenamefont {Ju}}]{zhao2023insight}%
  \BibitemOpen
  \bibfield  {author} {\bibinfo {author} {\bibfnamefont {R.}~\bibnamefont
  {Zhao}}, \bibinfo {author} {\bibfnamefont {C.}~\bibnamefont {Wan}}, \bibinfo
  {author} {\bibfnamefont {P.}~\bibnamefont {Qian}},\ and\ \bibinfo {author}
  {\bibfnamefont {X.}~\bibnamefont {Ju}},\ }\href@noop {} {\bibfield  {journal}
  {\bibinfo  {journal} {Surfaces and Interfaces}\ }\textbf {\bibinfo {volume}
  {38}},\ \bibinfo {pages} {102851} (\bibinfo {year}
  {2023}{\natexlab{b}})}\BibitemShut {NoStop}%
\bibitem [{\citenamefont {Cui}\ \emph {et~al.}(2020)\citenamefont {Cui},
  \citenamefont {Jia},\ and\ \citenamefont {Peng}}]{cui2020ad}%
  \BibitemOpen
  \bibfield  {author} {\bibinfo {author} {\bibfnamefont {H.}~\bibnamefont
  {Cui}}, \bibinfo {author} {\bibfnamefont {P.}~\bibnamefont {Jia}},\ and\
  \bibinfo {author} {\bibfnamefont {X.}~\bibnamefont {Peng}},\ }\href@noop {}
  {\bibfield  {journal} {\bibinfo  {journal} {Applied Surface Science}\
  }\textbf {\bibinfo {volume} {513}},\ \bibinfo {pages} {145863} (\bibinfo
  {year} {2020})}\BibitemShut {NoStop}%
\bibitem [{\citenamefont {Zhao}\ \emph
  {et~al.}(2023{\natexlab{c}})\citenamefont {Zhao}, \citenamefont {Tripathi},
  \citenamefont {{\v{C}}er{\c{n}}evi{\v{c}}s}, \citenamefont {Avsar},
  \citenamefont {Ji}, \citenamefont {Gonzalez~Marin}, \citenamefont {Cheon},
  \citenamefont {Wang}, \citenamefont {Yazyev},\ and\ \citenamefont
  {Kis}}]{zhao2023electrical}%
  \BibitemOpen
  \bibfield  {author} {\bibinfo {author} {\bibfnamefont {Y.}~\bibnamefont
  {Zhao}}, \bibinfo {author} {\bibfnamefont {M.}~\bibnamefont {Tripathi}},
  \bibinfo {author} {\bibfnamefont {K.}~\bibnamefont
  {{\v{C}}er{\c{n}}evi{\v{c}}s}}, \bibinfo {author} {\bibfnamefont
  {A.}~\bibnamefont {Avsar}}, \bibinfo {author} {\bibfnamefont {H.~G.}\
  \bibnamefont {Ji}}, \bibinfo {author} {\bibfnamefont {J.~F.}\ \bibnamefont
  {Gonzalez~Marin}}, \bibinfo {author} {\bibfnamefont {C.-Y.}\ \bibnamefont
  {Cheon}}, \bibinfo {author} {\bibfnamefont {Z.}~\bibnamefont {Wang}},
  \bibinfo {author} {\bibfnamefont {O.~V.}\ \bibnamefont {Yazyev}},\ and\
  \bibinfo {author} {\bibfnamefont {A.}~\bibnamefont {Kis}},\ }\href@noop {}
  {\bibfield  {journal} {\bibinfo  {journal} {Nature Communications}\ }\textbf
  {\bibinfo {volume} {14}},\ \bibinfo {pages} {44} (\bibinfo {year}
  {2023}{\natexlab{c}})}\BibitemShut {NoStop}%
\bibitem [{\citenamefont {Tarnopolsky}\ \emph {et~al.}(2019)\citenamefont
  {Tarnopolsky}, \citenamefont {Kruchkov},\ and\ \citenamefont
  {Vishwanath}}]{tarnopolsky2019origin}%
  \BibitemOpen
  \bibfield  {author} {\bibinfo {author} {\bibfnamefont {G.}~\bibnamefont
  {Tarnopolsky}}, \bibinfo {author} {\bibfnamefont {A.~J.}\ \bibnamefont
  {Kruchkov}},\ and\ \bibinfo {author} {\bibfnamefont {A.}~\bibnamefont
  {Vishwanath}},\ }\href@noop {} {\bibfield  {journal} {\bibinfo  {journal}
  {Physical review letters}\ }\textbf {\bibinfo {volume} {122}},\ \bibinfo
  {pages} {106405} (\bibinfo {year} {2019})}\BibitemShut {NoStop}%
\bibitem [{\citenamefont {Tilak}\ \emph {et~al.}(2021)\citenamefont {Tilak},
  \citenamefont {Lai}, \citenamefont {Wu}, \citenamefont {Zhang}, \citenamefont
  {Xu}, \citenamefont {Ribeiro}, \citenamefont {Canfield},\ and\ \citenamefont
  {Andrei}}]{tilak2021flat}%
  \BibitemOpen
  \bibfield  {author} {\bibinfo {author} {\bibfnamefont {N.}~\bibnamefont
  {Tilak}}, \bibinfo {author} {\bibfnamefont {X.}~\bibnamefont {Lai}}, \bibinfo
  {author} {\bibfnamefont {S.}~\bibnamefont {Wu}}, \bibinfo {author}
  {\bibfnamefont {Z.}~\bibnamefont {Zhang}}, \bibinfo {author} {\bibfnamefont
  {M.}~\bibnamefont {Xu}}, \bibinfo {author} {\bibfnamefont {R.~d.~A.}\
  \bibnamefont {Ribeiro}}, \bibinfo {author} {\bibfnamefont {P.~C.}\
  \bibnamefont {Canfield}},\ and\ \bibinfo {author} {\bibfnamefont {E.~Y.}\
  \bibnamefont {Andrei}},\ }\href@noop {} {\bibfield  {journal} {\bibinfo
  {journal} {Nature communications}\ }\textbf {\bibinfo {volume} {12}},\
  \bibinfo {pages} {4180} (\bibinfo {year} {2021})}\BibitemShut {NoStop}%
\bibitem [{\citenamefont {Utama}\ \emph {et~al.}(2021)\citenamefont {Utama},
  \citenamefont {Koch}, \citenamefont {Lee}, \citenamefont {Leconte},
  \citenamefont {Li}, \citenamefont {Zhao}, \citenamefont {Jiang},
  \citenamefont {Zhu}, \citenamefont {Watanabe}, \citenamefont {Taniguchi}
  \emph {et~al.}}]{utama2021visualization}%
  \BibitemOpen
  \bibfield  {author} {\bibinfo {author} {\bibfnamefont {M.~I.~B.}\
  \bibnamefont {Utama}}, \bibinfo {author} {\bibfnamefont {R.~J.}\ \bibnamefont
  {Koch}}, \bibinfo {author} {\bibfnamefont {K.}~\bibnamefont {Lee}}, \bibinfo
  {author} {\bibfnamefont {N.}~\bibnamefont {Leconte}}, \bibinfo {author}
  {\bibfnamefont {H.}~\bibnamefont {Li}}, \bibinfo {author} {\bibfnamefont
  {S.}~\bibnamefont {Zhao}}, \bibinfo {author} {\bibfnamefont {L.}~\bibnamefont
  {Jiang}}, \bibinfo {author} {\bibfnamefont {J.}~\bibnamefont {Zhu}}, \bibinfo
  {author} {\bibfnamefont {K.}~\bibnamefont {Watanabe}}, \bibinfo {author}
  {\bibfnamefont {T.}~\bibnamefont {Taniguchi}}, \emph {et~al.},\ }\href@noop
  {} {\bibfield  {journal} {\bibinfo  {journal} {Nature Physics}\ }\textbf
  {\bibinfo {volume} {17}},\ \bibinfo {pages} {184} (\bibinfo {year}
  {2021})}\BibitemShut {NoStop}%
\bibitem [{\citenamefont {Jiang}\ \emph {et~al.}(2019)\citenamefont {Jiang},
  \citenamefont {Lai}, \citenamefont {Watanabe}, \citenamefont {Taniguchi},
  \citenamefont {Haule}, \citenamefont {Mao},\ and\ \citenamefont
  {Andrei}}]{jiang2019charge}%
  \BibitemOpen
  \bibfield  {author} {\bibinfo {author} {\bibfnamefont {Y.}~\bibnamefont
  {Jiang}}, \bibinfo {author} {\bibfnamefont {X.}~\bibnamefont {Lai}}, \bibinfo
  {author} {\bibfnamefont {K.}~\bibnamefont {Watanabe}}, \bibinfo {author}
  {\bibfnamefont {T.}~\bibnamefont {Taniguchi}}, \bibinfo {author}
  {\bibfnamefont {K.}~\bibnamefont {Haule}}, \bibinfo {author} {\bibfnamefont
  {J.}~\bibnamefont {Mao}},\ and\ \bibinfo {author} {\bibfnamefont {E.~Y.}\
  \bibnamefont {Andrei}},\ }\href@noop {} {\bibfield  {journal} {\bibinfo
  {journal} {Nature}\ }\textbf {\bibinfo {volume} {573}},\ \bibinfo {pages}
  {91} (\bibinfo {year} {2019})}\BibitemShut {NoStop}%
\bibitem [{\citenamefont {Yankowitz}\ \emph {et~al.}(2019)\citenamefont
  {Yankowitz}, \citenamefont {Chen}, \citenamefont {Polshyn}, \citenamefont
  {Zhang}, \citenamefont {Watanabe}, \citenamefont {Taniguchi}, \citenamefont
  {Graf}, \citenamefont {Young},\ and\ \citenamefont
  {Dean}}]{yankowitz2019tuning}%
  \BibitemOpen
  \bibfield  {author} {\bibinfo {author} {\bibfnamefont {M.}~\bibnamefont
  {Yankowitz}}, \bibinfo {author} {\bibfnamefont {S.}~\bibnamefont {Chen}},
  \bibinfo {author} {\bibfnamefont {H.}~\bibnamefont {Polshyn}}, \bibinfo
  {author} {\bibfnamefont {Y.}~\bibnamefont {Zhang}}, \bibinfo {author}
  {\bibfnamefont {K.}~\bibnamefont {Watanabe}}, \bibinfo {author}
  {\bibfnamefont {T.}~\bibnamefont {Taniguchi}}, \bibinfo {author}
  {\bibfnamefont {D.}~\bibnamefont {Graf}}, \bibinfo {author} {\bibfnamefont
  {A.~F.}\ \bibnamefont {Young}},\ and\ \bibinfo {author} {\bibfnamefont
  {C.~R.}\ \bibnamefont {Dean}},\ }\href@noop {} {\bibfield  {journal}
  {\bibinfo  {journal} {Science}\ }\textbf {\bibinfo {volume} {363}},\ \bibinfo
  {pages} {1059} (\bibinfo {year} {2019})}\BibitemShut {NoStop}%
\bibitem [{\citenamefont {Cao}\ \emph {et~al.}(2018{\natexlab{a}})\citenamefont
  {Cao}, \citenamefont {Fatemi}, \citenamefont {Fang}, \citenamefont
  {Watanabe}, \citenamefont {Taniguchi}, \citenamefont {Kaxiras},\ and\
  \citenamefont {Jarillo-Herrero}}]{cao2018unconventional}%
  \BibitemOpen
  \bibfield  {author} {\bibinfo {author} {\bibfnamefont {Y.}~\bibnamefont
  {Cao}}, \bibinfo {author} {\bibfnamefont {V.}~\bibnamefont {Fatemi}},
  \bibinfo {author} {\bibfnamefont {S.}~\bibnamefont {Fang}}, \bibinfo {author}
  {\bibfnamefont {K.}~\bibnamefont {Watanabe}}, \bibinfo {author}
  {\bibfnamefont {T.}~\bibnamefont {Taniguchi}}, \bibinfo {author}
  {\bibfnamefont {E.}~\bibnamefont {Kaxiras}},\ and\ \bibinfo {author}
  {\bibfnamefont {P.}~\bibnamefont {Jarillo-Herrero}},\ }\href@noop {}
  {\bibfield  {journal} {\bibinfo  {journal} {Nature}\ }\textbf {\bibinfo
  {volume} {556}},\ \bibinfo {pages} {43} (\bibinfo {year}
  {2018}{\natexlab{a}})}\BibitemShut {NoStop}%
\bibitem [{\citenamefont {Lu}\ \emph {et~al.}(2019)\citenamefont {Lu},
  \citenamefont {Stepanov}, \citenamefont {Yang}, \citenamefont {Xie},
  \citenamefont {Aamir}, \citenamefont {Das}, \citenamefont {Urgell},
  \citenamefont {Watanabe}, \citenamefont {Taniguchi}, \citenamefont {Zhang}
  \emph {et~al.}}]{lu2019superconductors}%
  \BibitemOpen
  \bibfield  {author} {\bibinfo {author} {\bibfnamefont {X.}~\bibnamefont
  {Lu}}, \bibinfo {author} {\bibfnamefont {P.}~\bibnamefont {Stepanov}},
  \bibinfo {author} {\bibfnamefont {W.}~\bibnamefont {Yang}}, \bibinfo {author}
  {\bibfnamefont {M.}~\bibnamefont {Xie}}, \bibinfo {author} {\bibfnamefont
  {M.~A.}\ \bibnamefont {Aamir}}, \bibinfo {author} {\bibfnamefont
  {I.}~\bibnamefont {Das}}, \bibinfo {author} {\bibfnamefont {C.}~\bibnamefont
  {Urgell}}, \bibinfo {author} {\bibfnamefont {K.}~\bibnamefont {Watanabe}},
  \bibinfo {author} {\bibfnamefont {T.}~\bibnamefont {Taniguchi}}, \bibinfo
  {author} {\bibfnamefont {G.}~\bibnamefont {Zhang}}, \emph {et~al.},\
  }\href@noop {} {\bibfield  {journal} {\bibinfo  {journal} {Nature}\ }\textbf
  {\bibinfo {volume} {574}},\ \bibinfo {pages} {653} (\bibinfo {year}
  {2019})}\BibitemShut {NoStop}%
\bibitem [{\citenamefont {Pons}\ \emph {et~al.}(2020)\citenamefont {Pons},
  \citenamefont {Mielke},\ and\ \citenamefont {Stauber}}]{pons2020flat}%
  \BibitemOpen
  \bibfield  {author} {\bibinfo {author} {\bibfnamefont {R.}~\bibnamefont
  {Pons}}, \bibinfo {author} {\bibfnamefont {A.}~\bibnamefont {Mielke}},\ and\
  \bibinfo {author} {\bibfnamefont {T.}~\bibnamefont {Stauber}},\ }\href@noop
  {} {\bibfield  {journal} {\bibinfo  {journal} {Physical Review B}\ }\textbf
  {\bibinfo {volume} {102}},\ \bibinfo {pages} {235101} (\bibinfo {year}
  {2020})}\BibitemShut {NoStop}%
\bibitem [{\citenamefont {Song}\ and\ \citenamefont
  {Bernevig}(2022)}]{song2022magic}%
  \BibitemOpen
  \bibfield  {author} {\bibinfo {author} {\bibfnamefont {Z.-D.}\ \bibnamefont
  {Song}}\ and\ \bibinfo {author} {\bibfnamefont {B.~A.}\ \bibnamefont
  {Bernevig}},\ }\href@noop {} {\bibfield  {journal} {\bibinfo  {journal}
  {Physical review letters}\ }\textbf {\bibinfo {volume} {129}},\ \bibinfo
  {pages} {047601} (\bibinfo {year} {2022})}\BibitemShut {NoStop}%
\bibitem [{\citenamefont {Liu}\ \emph {et~al.}(2020)\citenamefont {Liu},
  \citenamefont {Hao}, \citenamefont {Khalaf}, \citenamefont {Lee},
  \citenamefont {Ronen}, \citenamefont {Yoo}, \citenamefont {Haei~Najafabadi},
  \citenamefont {Watanabe}, \citenamefont {Taniguchi}, \citenamefont
  {Vishwanath} \emph {et~al.}}]{liu2020tunable}%
  \BibitemOpen
  \bibfield  {author} {\bibinfo {author} {\bibfnamefont {X.}~\bibnamefont
  {Liu}}, \bibinfo {author} {\bibfnamefont {Z.}~\bibnamefont {Hao}}, \bibinfo
  {author} {\bibfnamefont {E.}~\bibnamefont {Khalaf}}, \bibinfo {author}
  {\bibfnamefont {J.~Y.}\ \bibnamefont {Lee}}, \bibinfo {author} {\bibfnamefont
  {Y.}~\bibnamefont {Ronen}}, \bibinfo {author} {\bibfnamefont
  {H.}~\bibnamefont {Yoo}}, \bibinfo {author} {\bibfnamefont {D.}~\bibnamefont
  {Haei~Najafabadi}}, \bibinfo {author} {\bibfnamefont {K.}~\bibnamefont
  {Watanabe}}, \bibinfo {author} {\bibfnamefont {T.}~\bibnamefont {Taniguchi}},
  \bibinfo {author} {\bibfnamefont {A.}~\bibnamefont {Vishwanath}}, \emph
  {et~al.},\ }\href@noop {} {\bibfield  {journal} {\bibinfo  {journal}
  {Nature}\ }\textbf {\bibinfo {volume} {583}},\ \bibinfo {pages} {221}
  (\bibinfo {year} {2020})}\BibitemShut {NoStop}%
\bibitem [{\citenamefont {Han}\ \emph {et~al.}(2021)\citenamefont {Han},
  \citenamefont {Inoue}, \citenamefont {Fang}, \citenamefont {John},
  \citenamefont {Ye}, \citenamefont {Chan}, \citenamefont {Graf}, \citenamefont
  {Suzuki}, \citenamefont {Ghimire}, \citenamefont {Cho} \emph
  {et~al.}}]{han2021evidence}%
  \BibitemOpen
  \bibfield  {author} {\bibinfo {author} {\bibfnamefont {M.}~\bibnamefont
  {Han}}, \bibinfo {author} {\bibfnamefont {H.}~\bibnamefont {Inoue}}, \bibinfo
  {author} {\bibfnamefont {S.}~\bibnamefont {Fang}}, \bibinfo {author}
  {\bibfnamefont {C.}~\bibnamefont {John}}, \bibinfo {author} {\bibfnamefont
  {L.}~\bibnamefont {Ye}}, \bibinfo {author} {\bibfnamefont {M.~K.}\
  \bibnamefont {Chan}}, \bibinfo {author} {\bibfnamefont {D.}~\bibnamefont
  {Graf}}, \bibinfo {author} {\bibfnamefont {T.}~\bibnamefont {Suzuki}},
  \bibinfo {author} {\bibfnamefont {M.~P.}\ \bibnamefont {Ghimire}}, \bibinfo
  {author} {\bibfnamefont {W.~J.}\ \bibnamefont {Cho}}, \emph {et~al.},\
  }\href@noop {} {\bibfield  {journal} {\bibinfo  {journal} {Nature
  communications}\ }\textbf {\bibinfo {volume} {12}},\ \bibinfo {pages} {5345}
  (\bibinfo {year} {2021})}\BibitemShut {NoStop}%
\bibitem [{\citenamefont {Sun}\ \emph {et~al.}(2022)\citenamefont {Sun},
  \citenamefont {Zhou}, \citenamefont {Wang}, \citenamefont {Kumar},
  \citenamefont {Geng}, \citenamefont {Yue}, \citenamefont {Han}, \citenamefont
  {Haraguchi}, \citenamefont {Shimada}, \citenamefont {Cheng} \emph
  {et~al.}}]{sun2022observation}%
  \BibitemOpen
  \bibfield  {author} {\bibinfo {author} {\bibfnamefont {Z.}~\bibnamefont
  {Sun}}, \bibinfo {author} {\bibfnamefont {H.}~\bibnamefont {Zhou}}, \bibinfo
  {author} {\bibfnamefont {C.}~\bibnamefont {Wang}}, \bibinfo {author}
  {\bibfnamefont {S.}~\bibnamefont {Kumar}}, \bibinfo {author} {\bibfnamefont
  {D.}~\bibnamefont {Geng}}, \bibinfo {author} {\bibfnamefont {S.}~\bibnamefont
  {Yue}}, \bibinfo {author} {\bibfnamefont {X.}~\bibnamefont {Han}}, \bibinfo
  {author} {\bibfnamefont {Y.}~\bibnamefont {Haraguchi}}, \bibinfo {author}
  {\bibfnamefont {K.}~\bibnamefont {Shimada}}, \bibinfo {author} {\bibfnamefont
  {P.}~\bibnamefont {Cheng}}, \emph {et~al.},\ }\href@noop {} {\bibfield
  {journal} {\bibinfo  {journal} {Nano Letters}\ }\textbf {\bibinfo {volume}
  {22}},\ \bibinfo {pages} {4596} (\bibinfo {year} {2022})}\BibitemShut
  {NoStop}%
\bibitem [{\citenamefont {Kang}\ \emph {et~al.}(2020)\citenamefont {Kang},
  \citenamefont {Fang}, \citenamefont {Ye}, \citenamefont {Po}, \citenamefont
  {Denlinger}, \citenamefont {Jozwiak}, \citenamefont {Bostwick}, \citenamefont
  {Rotenberg}, \citenamefont {Kaxiras}, \citenamefont {Checkelsky} \emph
  {et~al.}}]{kang2020topological}%
  \BibitemOpen
  \bibfield  {author} {\bibinfo {author} {\bibfnamefont {M.}~\bibnamefont
  {Kang}}, \bibinfo {author} {\bibfnamefont {S.}~\bibnamefont {Fang}}, \bibinfo
  {author} {\bibfnamefont {L.}~\bibnamefont {Ye}}, \bibinfo {author}
  {\bibfnamefont {H.~C.}\ \bibnamefont {Po}}, \bibinfo {author} {\bibfnamefont
  {J.}~\bibnamefont {Denlinger}}, \bibinfo {author} {\bibfnamefont
  {C.}~\bibnamefont {Jozwiak}}, \bibinfo {author} {\bibfnamefont
  {A.}~\bibnamefont {Bostwick}}, \bibinfo {author} {\bibfnamefont
  {E.}~\bibnamefont {Rotenberg}}, \bibinfo {author} {\bibfnamefont
  {E.}~\bibnamefont {Kaxiras}}, \bibinfo {author} {\bibfnamefont {J.~G.}\
  \bibnamefont {Checkelsky}}, \emph {et~al.},\ }\href@noop {} {\bibfield
  {journal} {\bibinfo  {journal} {Nature communications}\ }\textbf {\bibinfo
  {volume} {11}},\ \bibinfo {pages} {4004} (\bibinfo {year}
  {2020})}\BibitemShut {NoStop}%
\bibitem [{\citenamefont {Guo}\ \emph {et~al.}(2023)\citenamefont {Guo},
  \citenamefont {Qi}, \citenamefont {Zhang}, \citenamefont {Gao}, \citenamefont
  {Hu}, \citenamefont {Zhou}, \citenamefont {Zang}, \citenamefont {Zhao},
  \citenamefont {Wang}, \citenamefont {Yan} \emph {et~al.}}]{guo2023ultrathin}%
  \BibitemOpen
  \bibfield  {author} {\bibinfo {author} {\bibfnamefont {Q.}~\bibnamefont
  {Guo}}, \bibinfo {author} {\bibfnamefont {X.-Z.}\ \bibnamefont {Qi}},
  \bibinfo {author} {\bibfnamefont {L.}~\bibnamefont {Zhang}}, \bibinfo
  {author} {\bibfnamefont {M.}~\bibnamefont {Gao}}, \bibinfo {author}
  {\bibfnamefont {S.}~\bibnamefont {Hu}}, \bibinfo {author} {\bibfnamefont
  {W.}~\bibnamefont {Zhou}}, \bibinfo {author} {\bibfnamefont {W.}~\bibnamefont
  {Zang}}, \bibinfo {author} {\bibfnamefont {X.}~\bibnamefont {Zhao}}, \bibinfo
  {author} {\bibfnamefont {J.}~\bibnamefont {Wang}}, \bibinfo {author}
  {\bibfnamefont {B.}~\bibnamefont {Yan}}, \emph {et~al.},\ }\href@noop {}
  {\bibfield  {journal} {\bibinfo  {journal} {Nature}\ }\textbf {\bibinfo
  {volume} {613}},\ \bibinfo {pages} {53} (\bibinfo {year} {2023})}\BibitemShut
  {NoStop}%
\bibitem [{\citenamefont {Mortazavi}\ \emph {et~al.}(2022)\citenamefont
  {Mortazavi}, \citenamefont {Shahrokhi}, \citenamefont {Javvaji},
  \citenamefont {Shapeev},\ and\ \citenamefont {Zhuang}}]{mortazavi2022highly}%
  \BibitemOpen
  \bibfield  {author} {\bibinfo {author} {\bibfnamefont {B.}~\bibnamefont
  {Mortazavi}}, \bibinfo {author} {\bibfnamefont {M.}~\bibnamefont
  {Shahrokhi}}, \bibinfo {author} {\bibfnamefont {B.}~\bibnamefont {Javvaji}},
  \bibinfo {author} {\bibfnamefont {A.~V.}\ \bibnamefont {Shapeev}},\ and\
  \bibinfo {author} {\bibfnamefont {X.}~\bibnamefont {Zhuang}},\ }\href@noop {}
  {\bibfield  {journal} {\bibinfo  {journal} {Nanotechnology}\ }\textbf
  {\bibinfo {volume} {33}},\ \bibinfo {pages} {275701} (\bibinfo {year}
  {2022})}\BibitemShut {NoStop}%
\bibitem [{\citenamefont {Jia}\ \emph {et~al.}(2019)\citenamefont {Jia},
  \citenamefont {Zhao}, \citenamefont {Gou}, \citenamefont {Zeng},\ and\
  \citenamefont {Li}}]{jia2019niobium}%
  \BibitemOpen
  \bibfield  {author} {\bibinfo {author} {\bibfnamefont {Y.}~\bibnamefont
  {Jia}}, \bibinfo {author} {\bibfnamefont {M.}~\bibnamefont {Zhao}}, \bibinfo
  {author} {\bibfnamefont {G.}~\bibnamefont {Gou}}, \bibinfo {author}
  {\bibfnamefont {X.~C.}\ \bibnamefont {Zeng}},\ and\ \bibinfo {author}
  {\bibfnamefont {J.}~\bibnamefont {Li}},\ }\href@noop {} {\bibfield  {journal}
  {\bibinfo  {journal} {Nanoscale Horizons}\ }\textbf {\bibinfo {volume} {4}},\
  \bibinfo {pages} {1113} (\bibinfo {year} {2019})}\BibitemShut {NoStop}%
\bibitem [{\citenamefont {Kresse}\ and\ \citenamefont
  {Furthm{\"u}ller}(1996)}]{kresse1996efficient}%
  \BibitemOpen
  \bibfield  {author} {\bibinfo {author} {\bibfnamefont {G.}~\bibnamefont
  {Kresse}}\ and\ \bibinfo {author} {\bibfnamefont {J.}~\bibnamefont
  {Furthm{\"u}ller}},\ }\href@noop {} {\bibfield  {journal} {\bibinfo
  {journal} {Physical review B}\ }\textbf {\bibinfo {volume} {54}},\ \bibinfo
  {pages} {11169} (\bibinfo {year} {1996})}\BibitemShut {NoStop}%
\bibitem [{\citenamefont {Perdew}\ \emph {et~al.}(1996)\citenamefont {Perdew},
  \citenamefont {Burke},\ and\ \citenamefont
  {Ernzerhof}}]{perdew1996generalized}%
  \BibitemOpen
  \bibfield  {author} {\bibinfo {author} {\bibfnamefont {J.~P.}\ \bibnamefont
  {Perdew}}, \bibinfo {author} {\bibfnamefont {K.}~\bibnamefont {Burke}},\ and\
  \bibinfo {author} {\bibfnamefont {M.}~\bibnamefont {Ernzerhof}},\ }\href@noop
  {} {\bibfield  {journal} {\bibinfo  {journal} {Physical review letters}\
  }\textbf {\bibinfo {volume} {77}},\ \bibinfo {pages} {3865} (\bibinfo {year}
  {1996})}\BibitemShut {NoStop}%
\bibitem [{\citenamefont {Gajdo{\v{s}}}\ \emph {et~al.}(2006)\citenamefont
  {Gajdo{\v{s}}}, \citenamefont {Hummer}, \citenamefont {Kresse}, \citenamefont
  {Furthm{\"u}ller},\ and\ \citenamefont {Bechstedt}}]{gajdovs2006linear}%
  \BibitemOpen
  \bibfield  {author} {\bibinfo {author} {\bibfnamefont {M.}~\bibnamefont
  {Gajdo{\v{s}}}}, \bibinfo {author} {\bibfnamefont {K.}~\bibnamefont
  {Hummer}}, \bibinfo {author} {\bibfnamefont {G.}~\bibnamefont {Kresse}},
  \bibinfo {author} {\bibfnamefont {J.}~\bibnamefont {Furthm{\"u}ller}},\ and\
  \bibinfo {author} {\bibfnamefont {F.}~\bibnamefont {Bechstedt}},\ }\href@noop
  {} {\bibfield  {journal} {\bibinfo  {journal} {Physical Review B}\ }\textbf
  {\bibinfo {volume} {73}},\ \bibinfo {pages} {045112} (\bibinfo {year}
  {2006})}\BibitemShut {NoStop}%
\bibitem [{\citenamefont {Grimme}(2006)}]{grimme2006s}%
  \BibitemOpen
  \bibfield  {author} {\bibinfo {author} {\bibfnamefont {S.}~\bibnamefont
  {Grimme}},\ }\href@noop {} {\bibfield  {journal} {\bibinfo  {journal}
  {Journal of computational chemistry}\ }\textbf {\bibinfo {volume} {27}},\
  \bibinfo {pages} {1787} (\bibinfo {year} {2006})}\BibitemShut {NoStop}%
\bibitem [{\citenamefont {Peralta}\ \emph {et~al.}(2006)\citenamefont
  {Peralta}, \citenamefont {Heyd}, \citenamefont {Scuseria},\ and\
  \citenamefont {Martin}}]{peralta2006spin}%
  \BibitemOpen
  \bibfield  {author} {\bibinfo {author} {\bibfnamefont {J.~E.}\ \bibnamefont
  {Peralta}}, \bibinfo {author} {\bibfnamefont {J.}~\bibnamefont {Heyd}},
  \bibinfo {author} {\bibfnamefont {G.~E.}\ \bibnamefont {Scuseria}},\ and\
  \bibinfo {author} {\bibfnamefont {R.~L.}\ \bibnamefont {Martin}},\
  }\href@noop {} {\bibfield  {journal} {\bibinfo  {journal} {Physical Review
  B}\ }\textbf {\bibinfo {volume} {74}},\ \bibinfo {pages} {073101} (\bibinfo
  {year} {2006})}\BibitemShut {NoStop}%
\bibitem [{\citenamefont {Henkelman}\ \emph {et~al.}(2006)\citenamefont
  {Henkelman}, \citenamefont {Arnaldsson},\ and\ \citenamefont
  {J{\'o}nsson}}]{henkelman2006fast}%
  \BibitemOpen
  \bibfield  {author} {\bibinfo {author} {\bibfnamefont {G.}~\bibnamefont
  {Henkelman}}, \bibinfo {author} {\bibfnamefont {A.}~\bibnamefont
  {Arnaldsson}},\ and\ \bibinfo {author} {\bibfnamefont {H.}~\bibnamefont
  {J{\'o}nsson}},\ }\href@noop {} {\bibfield  {journal} {\bibinfo  {journal}
  {Computational Materials Science}\ }\textbf {\bibinfo {volume} {36}},\
  \bibinfo {pages} {354} (\bibinfo {year} {2006})}\BibitemShut {NoStop}%
\bibitem [{\citenamefont {Hybertsen}\ and\ \citenamefont
  {Louie}(1986)}]{hybertsen1986}%
  \BibitemOpen
  \bibfield  {author} {\bibinfo {author} {\bibfnamefont {M.~S.}\ \bibnamefont
  {Hybertsen}}\ and\ \bibinfo {author} {\bibfnamefont {S.~G.}\ \bibnamefont
  {Louie}},\ }\href@noop {} {\bibfield  {journal} {\bibinfo  {journal}
  {Physical Review B}\ }\textbf {\bibinfo {volume} {34}},\ \bibinfo {pages}
  {5390} (\bibinfo {year} {1986})}\BibitemShut {NoStop}%
\bibitem [{\citenamefont {Mohebpour}\ \emph
  {et~al.}(2022{\natexlab{b}})\citenamefont {Mohebpour}, \citenamefont
  {Mozvashi}, \citenamefont {Vishkayi},\ and\ \citenamefont
  {Tagani}}]{mohebpour2022tra}%
  \BibitemOpen
  \bibfield  {author} {\bibinfo {author} {\bibfnamefont {M.~A.}\ \bibnamefont
  {Mohebpour}}, \bibinfo {author} {\bibfnamefont {S.~M.}\ \bibnamefont
  {Mozvashi}}, \bibinfo {author} {\bibfnamefont {S.~I.}\ \bibnamefont
  {Vishkayi}},\ and\ \bibinfo {author} {\bibfnamefont {M.~B.}\ \bibnamefont
  {Tagani}},\ }\href@noop {} {\bibfield  {journal} {\bibinfo  {journal}
  {Physical Review Materials}\ }\textbf {\bibinfo {volume} {6}},\ \bibinfo
  {pages} {014012} (\bibinfo {year} {2022}{\natexlab{b}})}\BibitemShut
  {NoStop}%
\bibitem [{\citenamefont {Onida}\ \emph {et~al.}(2002)\citenamefont {Onida},
  \citenamefont {Reining},\ and\ \citenamefont {Rubio}}]{onida2002elc}%
  \BibitemOpen
  \bibfield  {author} {\bibinfo {author} {\bibfnamefont {G.}~\bibnamefont
  {Onida}}, \bibinfo {author} {\bibfnamefont {L.}~\bibnamefont {Reining}},\
  and\ \bibinfo {author} {\bibfnamefont {A.}~\bibnamefont {Rubio}},\
  }\href@noop {} {\bibfield  {journal} {\bibinfo  {journal} {Reviews of modern
  physics}\ }\textbf {\bibinfo {volume} {74}},\ \bibinfo {pages} {601}
  (\bibinfo {year} {2002})}\BibitemShut {NoStop}%
\bibitem [{\citenamefont {Togo}\ \emph {et~al.}(2008)\citenamefont {Togo},
  \citenamefont {Oba},\ and\ \citenamefont {Tanaka}}]{togo2008}%
  \BibitemOpen
  \bibfield  {author} {\bibinfo {author} {\bibfnamefont {A.}~\bibnamefont
  {Togo}}, \bibinfo {author} {\bibfnamefont {F.}~\bibnamefont {Oba}},\ and\
  \bibinfo {author} {\bibfnamefont {I.}~\bibnamefont {Tanaka}},\ }\href@noop {}
  {\bibfield  {journal} {\bibinfo  {journal} {Physical Review B}\ }\textbf
  {\bibinfo {volume} {78}},\ \bibinfo {pages} {134106} (\bibinfo {year}
  {2008})}\BibitemShut {NoStop}%
\bibitem [{\citenamefont {Mohebpour}\ \emph
  {et~al.}(2020{\natexlab{a}})\citenamefont {Mohebpour}, \citenamefont
  {Vishkayi},\ and\ \citenamefont {Tagani}}]{mohetuning}%
  \BibitemOpen
  \bibfield  {author} {\bibinfo {author} {\bibfnamefont {M.~A.}\ \bibnamefont
  {Mohebpour}}, \bibinfo {author} {\bibfnamefont {S.~I.}\ \bibnamefont
  {Vishkayi}},\ and\ \bibinfo {author} {\bibfnamefont {M.~B.}\ \bibnamefont
  {Tagani}},\ }\href@noop {} {\bibfield  {journal} {\bibinfo  {journal}
  {Journal of Applied Physics}\ }\textbf {\bibinfo {volume} {127}} (\bibinfo
  {year} {2020}{\natexlab{a}})}\BibitemShut {NoStop}%
\bibitem [{\citenamefont {Mohebpour}\ \emph
  {et~al.}(2020{\natexlab{b}})\citenamefont {Mohebpour}, \citenamefont
  {Mozvashi}, \citenamefont {Vishkayi},\ and\ \citenamefont
  {Tagani}}]{mohebpourtion}%
  \BibitemOpen
  \bibfield  {author} {\bibinfo {author} {\bibfnamefont {M.~A.}\ \bibnamefont
  {Mohebpour}}, \bibinfo {author} {\bibfnamefont {S.~M.}\ \bibnamefont
  {Mozvashi}}, \bibinfo {author} {\bibfnamefont {S.~I.}\ \bibnamefont
  {Vishkayi}},\ and\ \bibinfo {author} {\bibfnamefont {M.~B.}\ \bibnamefont
  {Tagani}},\ }\href@noop {} {\bibfield  {journal} {\bibinfo  {journal}
  {Scientific reports}\ }\textbf {\bibinfo {volume} {10}},\ \bibinfo {pages}
  {14963} (\bibinfo {year} {2020}{\natexlab{b}})}\BibitemShut {NoStop}%
\bibitem [{\citenamefont {Wu}\ \emph {et~al.}(2022)\citenamefont {Wu},
  \citenamefont {Abdelwahab}, \citenamefont {Kwon}, \citenamefont
  {Verzhbitskiy}, \citenamefont {Wang}, \citenamefont {Liew}, \citenamefont
  {Yao}, \citenamefont {Eda}, \citenamefont {Loh}, \citenamefont {Shen} \emph
  {et~al.}}]{wu2022data}%
  \BibitemOpen
  \bibfield  {author} {\bibinfo {author} {\bibfnamefont {Y.}~\bibnamefont
  {Wu}}, \bibinfo {author} {\bibfnamefont {I.}~\bibnamefont {Abdelwahab}},
  \bibinfo {author} {\bibfnamefont {K.~C.}\ \bibnamefont {Kwon}}, \bibinfo
  {author} {\bibfnamefont {I.}~\bibnamefont {Verzhbitskiy}}, \bibinfo {author}
  {\bibfnamefont {L.}~\bibnamefont {Wang}}, \bibinfo {author} {\bibfnamefont
  {W.~H.}\ \bibnamefont {Liew}}, \bibinfo {author} {\bibfnamefont
  {K.}~\bibnamefont {Yao}}, \bibinfo {author} {\bibfnamefont {G.}~\bibnamefont
  {Eda}}, \bibinfo {author} {\bibfnamefont {K.~P.}\ \bibnamefont {Loh}},
  \bibinfo {author} {\bibfnamefont {L.}~\bibnamefont {Shen}}, \emph {et~al.},\
  }\href@noop {} {\bibfield  {journal} {\bibinfo  {journal} {Nature
  Communications}\ }\textbf {\bibinfo {volume} {13}},\ \bibinfo {pages} {1884}
  (\bibinfo {year} {2022})}\BibitemShut {NoStop}%
\bibitem [{\citenamefont {Chen}\ \emph
  {et~al.}(2022{\natexlab{b}})\citenamefont {Chen}, \citenamefont {Deng},
  \citenamefont {Kan}, \citenamefont {Yan}, \citenamefont {Shi}, \citenamefont
  {Huo}, \citenamefont {Song}, \citenamefont {Yang},\ and\ \citenamefont
  {Liu}}]{chen2022ferromagnetic}%
  \BibitemOpen
  \bibfield  {author} {\bibinfo {author} {\bibfnamefont {K.}~\bibnamefont
  {Chen}}, \bibinfo {author} {\bibfnamefont {J.}~\bibnamefont {Deng}}, \bibinfo
  {author} {\bibfnamefont {D.}~\bibnamefont {Kan}}, \bibinfo {author}
  {\bibfnamefont {Y.}~\bibnamefont {Yan}}, \bibinfo {author} {\bibfnamefont
  {Q.}~\bibnamefont {Shi}}, \bibinfo {author} {\bibfnamefont {W.}~\bibnamefont
  {Huo}}, \bibinfo {author} {\bibfnamefont {M.}~\bibnamefont {Song}}, \bibinfo
  {author} {\bibfnamefont {S.}~\bibnamefont {Yang}},\ and\ \bibinfo {author}
  {\bibfnamefont {J.~Z.}\ \bibnamefont {Liu}},\ }\href@noop {} {\bibfield
  {journal} {\bibinfo  {journal} {Physical Review B}\ }\textbf {\bibinfo
  {volume} {105}},\ \bibinfo {pages} {024414} (\bibinfo {year}
  {2022}{\natexlab{b}})}\BibitemShut {NoStop}%
\bibitem [{\citenamefont {Cudazzo}\ and\ \citenamefont
  {Wirtz}(2021)}]{cudazzo2021collective}%
  \BibitemOpen
  \bibfield  {author} {\bibinfo {author} {\bibfnamefont {P.}~\bibnamefont
  {Cudazzo}}\ and\ \bibinfo {author} {\bibfnamefont {L.}~\bibnamefont
  {Wirtz}},\ }\href@noop {} {\bibfield  {journal} {\bibinfo  {journal}
  {Physical Review B}\ }\textbf {\bibinfo {volume} {104}},\ \bibinfo {pages}
  {125101} (\bibinfo {year} {2021})}\BibitemShut {NoStop}%
\bibitem [{\citenamefont {Bhattacharya}\ \emph {et~al.}(2023)\citenamefont
  {Bhattacharya}, \citenamefont {Timokhin}, \citenamefont {Chatterjee},
  \citenamefont {Yang},\ and\ \citenamefont
  {Mishchenko}}]{bhattacharya2023deep}%
  \BibitemOpen
  \bibfield  {author} {\bibinfo {author} {\bibfnamefont {A.}~\bibnamefont
  {Bhattacharya}}, \bibinfo {author} {\bibfnamefont {I.}~\bibnamefont
  {Timokhin}}, \bibinfo {author} {\bibfnamefont {R.}~\bibnamefont
  {Chatterjee}}, \bibinfo {author} {\bibfnamefont {Q.}~\bibnamefont {Yang}},\
  and\ \bibinfo {author} {\bibfnamefont {A.}~\bibnamefont {Mishchenko}},\
  }\href@noop {} {\bibfield  {journal} {\bibinfo  {journal} {npj Computational
  Materials}\ }\textbf {\bibinfo {volume} {9}},\ \bibinfo {pages} {101}
  (\bibinfo {year} {2023})}\BibitemShut {NoStop}%
\bibitem [{\citenamefont {Li}\ \emph {et~al.}(2023)\citenamefont {Li},
  \citenamefont {Xue}, \citenamefont {Yao}, \citenamefont {Zhao}, \citenamefont
  {Xu},\ and\ \citenamefont {Yang}}]{li2023new}%
  \BibitemOpen
  \bibfield  {author} {\bibinfo {author} {\bibfnamefont {Z.}~\bibnamefont
  {Li}}, \bibinfo {author} {\bibfnamefont {Y.}~\bibnamefont {Xue}}, \bibinfo
  {author} {\bibfnamefont {Q.}~\bibnamefont {Yao}}, \bibinfo {author}
  {\bibfnamefont {B.}~\bibnamefont {Zhao}}, \bibinfo {author} {\bibfnamefont
  {W.}~\bibnamefont {Xu}},\ and\ \bibinfo {author} {\bibfnamefont
  {Z.}~\bibnamefont {Yang}},\ }\href@noop {} {\bibfield  {journal} {\bibinfo
  {journal} {Nanotechnology}\ } (\bibinfo {year} {2023})}\BibitemShut {NoStop}%
\bibitem [{\citenamefont {Yang}\ \emph {et~al.}(2022)\citenamefont {Yang},
  \citenamefont {Gou}, \citenamefont {Lu},\ and\ \citenamefont
  {Hao}}]{yang2022linear}%
  \BibitemOpen
  \bibfield  {author} {\bibinfo {author} {\bibfnamefont {N.}~\bibnamefont
  {Yang}}, \bibinfo {author} {\bibfnamefont {G.}~\bibnamefont {Gou}}, \bibinfo
  {author} {\bibfnamefont {X.}~\bibnamefont {Lu}},\ and\ \bibinfo {author}
  {\bibfnamefont {Y.}~\bibnamefont {Hao}},\ }\href@noop {} {\bibfield
  {journal} {\bibinfo  {journal} {Nano Research}\ }\textbf {\bibinfo {volume}
  {15}},\ \bibinfo {pages} {6779} (\bibinfo {year} {2022})}\BibitemShut
  {NoStop}%
\bibitem [{\citenamefont {Liu}\ \emph {et~al.}(2014)\citenamefont {Liu},
  \citenamefont {Liu},\ and\ \citenamefont {Wu}}]{liu2014exotic}%
  \BibitemOpen
  \bibfield  {author} {\bibinfo {author} {\bibfnamefont {Z.}~\bibnamefont
  {Liu}}, \bibinfo {author} {\bibfnamefont {F.}~\bibnamefont {Liu}},\ and\
  \bibinfo {author} {\bibfnamefont {Y.-S.}\ \bibnamefont {Wu}},\ }\href@noop {}
  {\bibfield  {journal} {\bibinfo  {journal} {Chinese Physics B}\ }\textbf
  {\bibinfo {volume} {23}},\ \bibinfo {pages} {077308} (\bibinfo {year}
  {2014})}\BibitemShut {NoStop}%
\bibitem [{\citenamefont {Zhang}\ \emph {et~al.}(2021)\citenamefont {Zhang},
  \citenamefont {Lin}, \citenamefont {Moreo}, \citenamefont {Alvarez},\ and\
  \citenamefont {Dagotto}}]{zhang2021peierls}%
  \BibitemOpen
  \bibfield  {author} {\bibinfo {author} {\bibfnamefont {Y.}~\bibnamefont
  {Zhang}}, \bibinfo {author} {\bibfnamefont {L.-F.}\ \bibnamefont {Lin}},
  \bibinfo {author} {\bibfnamefont {A.}~\bibnamefont {Moreo}}, \bibinfo
  {author} {\bibfnamefont {G.}~\bibnamefont {Alvarez}},\ and\ \bibinfo {author}
  {\bibfnamefont {E.}~\bibnamefont {Dagotto}},\ }\href@noop {} {\bibfield
  {journal} {\bibinfo  {journal} {Physical Review B}\ }\textbf {\bibinfo
  {volume} {103}},\ \bibinfo {pages} {L121114} (\bibinfo {year}
  {2021})}\BibitemShut {NoStop}%
\bibitem [{\citenamefont {Chowdhury}\ \emph {et~al.}(2017)\citenamefont
  {Chowdhury}, \citenamefont {Karmakar},\ and\ \citenamefont
  {Datta}}]{chowdhury2017monolayer}%
  \BibitemOpen
  \bibfield  {author} {\bibinfo {author} {\bibfnamefont {C.}~\bibnamefont
  {Chowdhury}}, \bibinfo {author} {\bibfnamefont {S.}~\bibnamefont
  {Karmakar}},\ and\ \bibinfo {author} {\bibfnamefont {A.}~\bibnamefont
  {Datta}},\ }\href@noop {} {\bibfield  {journal} {\bibinfo  {journal} {The
  Journal of Physical Chemistry C}\ }\textbf {\bibinfo {volume} {121}},\
  \bibinfo {pages} {7615} (\bibinfo {year} {2017})}\BibitemShut {NoStop}%
\bibitem [{\citenamefont {Tran}\ \emph {et~al.}(2014)\citenamefont {Tran},
  \citenamefont {Soklaski}, \citenamefont {Liang},\ and\ \citenamefont
  {Yang}}]{tran2014layer}%
  \BibitemOpen
  \bibfield  {author} {\bibinfo {author} {\bibfnamefont {V.}~\bibnamefont
  {Tran}}, \bibinfo {author} {\bibfnamefont {R.}~\bibnamefont {Soklaski}},
  \bibinfo {author} {\bibfnamefont {Y.}~\bibnamefont {Liang}},\ and\ \bibinfo
  {author} {\bibfnamefont {L.}~\bibnamefont {Yang}},\ }\href@noop {} {\bibfield
   {journal} {\bibinfo  {journal} {Physical Review B}\ }\textbf {\bibinfo
  {volume} {89}},\ \bibinfo {pages} {235319} (\bibinfo {year}
  {2014})}\BibitemShut {NoStop}%
\bibitem [{\citenamefont {Cao}\ \emph {et~al.}(2018{\natexlab{b}})\citenamefont
  {Cao}, \citenamefont {Wu},\ and\ \citenamefont {Louie}}]{cao2018unifying}%
  \BibitemOpen
  \bibfield  {author} {\bibinfo {author} {\bibfnamefont {T.}~\bibnamefont
  {Cao}}, \bibinfo {author} {\bibfnamefont {M.}~\bibnamefont {Wu}},\ and\
  \bibinfo {author} {\bibfnamefont {S.~G.}\ \bibnamefont {Louie}},\ }\href@noop
  {} {\bibfield  {journal} {\bibinfo  {journal} {Physical review letters}\
  }\textbf {\bibinfo {volume} {120}},\ \bibinfo {pages} {087402} (\bibinfo
  {year} {2018}{\natexlab{b}})}\BibitemShut {NoStop}%
\bibitem [{\citenamefont {Sethi}\ \emph {et~al.}(2021)\citenamefont {Sethi},
  \citenamefont {Zhou}, \citenamefont {Zhu}, \citenamefont {Yang},\ and\
  \citenamefont {Liu}}]{sethi2021flat}%
  \BibitemOpen
  \bibfield  {author} {\bibinfo {author} {\bibfnamefont {G.}~\bibnamefont
  {Sethi}}, \bibinfo {author} {\bibfnamefont {Y.}~\bibnamefont {Zhou}},
  \bibinfo {author} {\bibfnamefont {L.}~\bibnamefont {Zhu}}, \bibinfo {author}
  {\bibfnamefont {L.}~\bibnamefont {Yang}},\ and\ \bibinfo {author}
  {\bibfnamefont {F.}~\bibnamefont {Liu}},\ }\href@noop {} {\bibfield
  {journal} {\bibinfo  {journal} {Physical Review Letters}\ }\textbf {\bibinfo
  {volume} {126}},\ \bibinfo {pages} {196403} (\bibinfo {year}
  {2021})}\BibitemShut {NoStop}%
\bibitem [{\citenamefont {Deng}\ \emph {et~al.}(2018)\citenamefont {Deng},
  \citenamefont {Yu}, \citenamefont {Song}, \citenamefont {Zhang},
  \citenamefont {Wang}, \citenamefont {Sun}, \citenamefont {Yi}, \citenamefont
  {Wu}, \citenamefont {Wu}, \citenamefont {Zhu} \emph {et~al.}}]{deng2018gate}%
  \BibitemOpen
  \bibfield  {author} {\bibinfo {author} {\bibfnamefont {Y.}~\bibnamefont
  {Deng}}, \bibinfo {author} {\bibfnamefont {Y.}~\bibnamefont {Yu}}, \bibinfo
  {author} {\bibfnamefont {Y.}~\bibnamefont {Song}}, \bibinfo {author}
  {\bibfnamefont {J.}~\bibnamefont {Zhang}}, \bibinfo {author} {\bibfnamefont
  {N.~Z.}\ \bibnamefont {Wang}}, \bibinfo {author} {\bibfnamefont
  {Z.}~\bibnamefont {Sun}}, \bibinfo {author} {\bibfnamefont {Y.}~\bibnamefont
  {Yi}}, \bibinfo {author} {\bibfnamefont {Y.~Z.}\ \bibnamefont {Wu}}, \bibinfo
  {author} {\bibfnamefont {S.}~\bibnamefont {Wu}}, \bibinfo {author}
  {\bibfnamefont {J.}~\bibnamefont {Zhu}}, \emph {et~al.},\ }\href@noop {}
  {\bibfield  {journal} {\bibinfo  {journal} {Nature}\ }\textbf {\bibinfo
  {volume} {563}},\ \bibinfo {pages} {94} (\bibinfo {year} {2018})}\BibitemShut
  {NoStop}%
\bibitem [{\citenamefont {Wang}\ \emph {et~al.}(2018)\citenamefont {Wang},
  \citenamefont {Zhang}, \citenamefont {Ding}, \citenamefont {Dong},
  \citenamefont {Li}, \citenamefont {Chen}, \citenamefont {Li}, \citenamefont
  {Huang}, \citenamefont {Wang}, \citenamefont {Zhao} \emph
  {et~al.}}]{wang2018electric}%
  \BibitemOpen
  \bibfield  {author} {\bibinfo {author} {\bibfnamefont {Z.}~\bibnamefont
  {Wang}}, \bibinfo {author} {\bibfnamefont {T.}~\bibnamefont {Zhang}},
  \bibinfo {author} {\bibfnamefont {M.}~\bibnamefont {Ding}}, \bibinfo {author}
  {\bibfnamefont {B.}~\bibnamefont {Dong}}, \bibinfo {author} {\bibfnamefont
  {Y.}~\bibnamefont {Li}}, \bibinfo {author} {\bibfnamefont {M.}~\bibnamefont
  {Chen}}, \bibinfo {author} {\bibfnamefont {X.}~\bibnamefont {Li}}, \bibinfo
  {author} {\bibfnamefont {J.}~\bibnamefont {Huang}}, \bibinfo {author}
  {\bibfnamefont {H.}~\bibnamefont {Wang}}, \bibinfo {author} {\bibfnamefont
  {X.}~\bibnamefont {Zhao}}, \emph {et~al.},\ }\href@noop {} {\bibfield
  {journal} {\bibinfo  {journal} {Nature nanotechnology}\ }\textbf {\bibinfo
  {volume} {13}},\ \bibinfo {pages} {554} (\bibinfo {year} {2018})}\BibitemShut
  {NoStop}%
\bibitem [{\citenamefont {Zhang}\ and\ \citenamefont
  {Liu}(2018)}]{zhang2018hole}%
  \BibitemOpen
  \bibfield  {author} {\bibinfo {author} {\bibfnamefont {S.-H.}\ \bibnamefont
  {Zhang}}\ and\ \bibinfo {author} {\bibfnamefont {B.-G.}\ \bibnamefont
  {Liu}},\ }\href@noop {} {\bibfield  {journal} {\bibinfo  {journal} {Journal
  of Materials Chemistry C}\ }\textbf {\bibinfo {volume} {6}},\ \bibinfo
  {pages} {6792} (\bibinfo {year} {2018})}\BibitemShut {NoStop}%
\bibitem [{\citenamefont {Meng}\ \emph {et~al.}(2020)\citenamefont {Meng},
  \citenamefont {Houssa}, \citenamefont {Iordanidou}, \citenamefont {Pourtois},
  \citenamefont {Afanasiev},\ and\ \citenamefont {Stesmans}}]{meng2020ferro}%
  \BibitemOpen
  \bibfield  {author} {\bibinfo {author} {\bibfnamefont {R.}~\bibnamefont
  {Meng}}, \bibinfo {author} {\bibfnamefont {M.}~\bibnamefont {Houssa}},
  \bibinfo {author} {\bibfnamefont {K.}~\bibnamefont {Iordanidou}}, \bibinfo
  {author} {\bibfnamefont {G.}~\bibnamefont {Pourtois}}, \bibinfo {author}
  {\bibfnamefont {V.}~\bibnamefont {Afanasiev}},\ and\ \bibinfo {author}
  {\bibfnamefont {A.}~\bibnamefont {Stesmans}},\ }\href@noop {} {\bibfield
  {journal} {\bibinfo  {journal} {Physical Review Materials}\ }\textbf
  {\bibinfo {volume} {4}},\ \bibinfo {pages} {074001} (\bibinfo {year}
  {2020})}\BibitemShut {NoStop}%
\bibitem [{\citenamefont {You}\ \emph {et~al.}(2019)\citenamefont {You},
  \citenamefont {Gu},\ and\ \citenamefont {Su}}]{you2019flat}%
  \BibitemOpen
  \bibfield  {author} {\bibinfo {author} {\bibfnamefont {J.-Y.}\ \bibnamefont
  {You}}, \bibinfo {author} {\bibfnamefont {B.}~\bibnamefont {Gu}},\ and\
  \bibinfo {author} {\bibfnamefont {G.}~\bibnamefont {Su}},\ }\href@noop {}
  {\bibfield  {journal} {\bibinfo  {journal} {Scientific reports}\ }\textbf
  {\bibinfo {volume} {9}},\ \bibinfo {pages} {20116} (\bibinfo {year}
  {2019})}\BibitemShut {NoStop}%
\bibitem [{\citenamefont {Lu}\ \emph {et~al.}(2023)\citenamefont {Lu},
  \citenamefont {Guo}, \citenamefont {Cheng}, \citenamefont {Guo},
  \citenamefont {Wang}, \citenamefont {Deng}, \citenamefont {Bai},
  \citenamefont {Tian}, \citenamefont {Zhou}, \citenamefont {Shi} \emph
  {et~al.}}]{lu2023controllable}%
  \BibitemOpen
  \bibfield  {author} {\bibinfo {author} {\bibfnamefont {S.}~\bibnamefont
  {Lu}}, \bibinfo {author} {\bibfnamefont {D.}~\bibnamefont {Guo}}, \bibinfo
  {author} {\bibfnamefont {Z.}~\bibnamefont {Cheng}}, \bibinfo {author}
  {\bibfnamefont {Y.}~\bibnamefont {Guo}}, \bibinfo {author} {\bibfnamefont
  {C.}~\bibnamefont {Wang}}, \bibinfo {author} {\bibfnamefont {J.}~\bibnamefont
  {Deng}}, \bibinfo {author} {\bibfnamefont {Y.}~\bibnamefont {Bai}}, \bibinfo
  {author} {\bibfnamefont {C.}~\bibnamefont {Tian}}, \bibinfo {author}
  {\bibfnamefont {L.}~\bibnamefont {Zhou}}, \bibinfo {author} {\bibfnamefont
  {Y.}~\bibnamefont {Shi}}, \emph {et~al.},\ }\href@noop {} {\bibfield
  {journal} {\bibinfo  {journal} {Nature Communications}\ }\textbf {\bibinfo
  {volume} {14}},\ \bibinfo {pages} {2465} (\bibinfo {year}
  {2023})}\BibitemShut {NoStop}%
\bibitem [{\citenamefont {Lee}(2023)}]{lee2023possible}%
  \BibitemOpen
  \bibfield  {author} {\bibinfo {author} {\bibfnamefont {Y.~H.}\ \bibnamefont
  {Lee}},\ }\href@noop {} {\bibfield  {journal} {\bibinfo  {journal} {Science}\
  }\textbf {\bibinfo {volume} {382}},\ \bibinfo {pages} {l0823} (\bibinfo
  {year} {2023})}\BibitemShut {NoStop}%
\end{thebibliography}%
\end{document}